\documentclass[12pt]{article}
\usepackage{amssymb}
\usepackage{amsfonts}
\usepackage{amssymb}
\usepackage{caption}
\usepackage{subcaption}
\usepackage{amsthm}\theoremstyle{remark}
\def\inh{\vskip 0.075truein \noindent\hangindent=12 pt \hangafter=1}

\usepackage{amsthm}\theoremstyle{remark}
\usepackage{graphicx}

\newcommand{\bte}{\begin{quote}\begin{theorem}}
\newcommand{\ete}[1]{\label{#1}\end{theorem}\end{quote}}
\newcommand{\bcom}{\begin{quote}\end{quote}}
\newcommand{\bex}{\begin{quote}\begin{example}}
\newcommand{\eex}[1]{\label{#1}\end{example}\end{quote}}
\newcommand{\bcon}{\begin{quote}\begin{conclusion}}
\newcommand{\econ}[1]{\label{#1}\end{conclusion}\end{quote}}
\newcommand{\bdefi}{\begin{quote}\begin{definition}}
\newcommand{\edefi}[1]{\label{#1}\end{definition}\end{quote}}

\newcommand{\blem}{\begin{quote}\begin{lemma}}
\newcommand{\elem}[1]{\label{#1}\end{lemma}\end{quote}}

\newcommand{\bpr}{\begin{quote}\begin{problem}}
\newcommand{\epr}[1]{\label{#1}\end{problem}\end{quote}}

\newcommand{\f}{\frac}

\newcommand{\n}{\nonumber \\}
\newcommand{\inti}{\int_{-\infty}^\infty}
\newcommand{\beq}{\begin{eqnarray}}
\newcommand{\eeq}[1]{\label{#1}\end{eqnarray}}
\newcommand\eq[1]{(\ref{#1})}
\newcommand{\bfi}{\begin{figure}[24]}
\newcommand{\efi}[1]{\caption{\label{#1}}\end{figure}}
\newcommand\fig[1]{Fig.~\ref{#1}}
\newcommand{\res}{respectively}
\newcommand\gl{\left}
\newcommand\gr{\right}

\newcommand{\CA}{{\cal A}}

\newcommand{\CE}{{\cal E}}

\newcommand{\CP}{{\cal P}}

\newcommand{\Ga}{\alpha}

\newcommand{\Gd}{\delta}

\newcommand{\Gf}{\phi}

\newcommand{\Gl}{\lambda}
\newcommand{\Gn}{\eta}
\newcommand{\Gm}{\mu}

\newcommand{\Go}{\omega}

\newcommand{\GT}{\Theta}

\newcommand{\az}[1]{Sect.$\!$ \ref{#1}}
\newcommand\D{\,\mathrm{d}}
\newcommand\I{\mathrm{i}}
\newcommand\E{\mathrm{e}}
\newcommand{\bexe}{\begin{quote}\begin{exercise}\inh}
\newcommand{\eexe}[1]{\label{#1}\end{exercise}\end{quote}}

\usepackage{graphics,graphicx}
\usepackage{color}

\begin{document}
{\large
\title{How a dissimilar-chain system is splitting \\ Quasi-static, subsonic and supersonic regimes }

\author{Igor E. Berinskii and Leonid I. Slepyan*}
\date{\small{{\em School of Mechanical Engineering, Tel Aviv University\\
P.O. Box 39040, Ramat Aviv 69978 Tel Aviv, Israel}}\\
 }}

\maketitle

\vspace{0mm}\noindent
{\bf Abstract}
\noindent
We consider parallel splitting of a strip composed of {\em two different chains}. As a waveguide, the dissimilar-chain structure radically differs from the well-studied identical-chain system. It is characterized by three speeds of the long waves, $c_1$ and $c_2$  for the separate chains, and $c_+=\sqrt{(c_1^2+c_2^2)/2}$ for the intact strip where the chains are connected. Accordingly, there exist three ranges, the subsonic for both chains $(0, c_2)$ (we assume that $c_2<c_1$), the  intersonic $(c_2, c_+)$  and the supersonic, $(c_+,c_1)$. The speed in the latter range is supersonic for the intact strip and at the same time, it is subsonic for the separate higher-speed chain. This fact allows the splitting wave to propagate in the strip supersonically.

We derive steady-state analytical solutions and find that the splitting can propagate steadily only in two of these speed ranges, the subsonic and the supersonic, whereas the intersonic regime is forbidden. In the case of considerable difference in the chain stiffness, the lowest dynamic threshold corresponds to the supersonic regime. The peculiarity of the supersonic mode is that the supersonic energy delivery channel, being initially absent, is opening with the moving splitting point.

Based on the discrete and related continuous models we find which regime can be implemented depending on the structure parameters and loading conditions. The analysis allows us to represent the characteristics of such processes and to demonstrate strengths and weaknesses of different formulations, quasi-static, dynamic, discrete or continuous.

Analytical solutions for steady-state regimes are obtained and analyzed in detail. We find the force $-$ speed relations and show the difference between the static and dynamic thresholds. The parameters and energy of waves radiated by the propagating splitting are determined. We calculate the strain distribution ahead of the transition point and check whether the steady-state solutions are admissible.

\vspace{10mm}\noindent

\vspace{10mm}\noindent
* Corresponding author. E-mail address: leonid@eng.tau.ac.il (L.I. Slepyan).

\section{Introduction}
Fracture mechanics (and the theory of transitions under driving forces in a broader sense), founded by Griffith (1920) and Eshelby (1951, 1956), developed over decades in the framework of continuum mechanics. Novozhilov (1969a,b), apparently for the first time, noticed the role of the medium discreteness and associated instabilities in the fracture process.
Thomson et al. (1971) independently drew attention to the importance of these factors, while examining quasi-static splitting of a double-chain strip. In their paper, the energy transferred to lattice oscillations the process of splitting was estimated and the term ``lattice trapping" was introduced. In the first analytical solution for the crack dynamics in a two-dimensional lattice (Slepyan, 1981), the local-to-global energy release ratio was presented as a function of the crack speed. Thereby, the speed-dependent wave radiation energy was determined. We also refer to the books by Kunin (1975, 1982, 1983) on the microstructure in elasticity.

To date, analytical examination of the lattice fracture and phase transition is substantially developed. Numerical methods and results in the molecular dynamics of lattices are described, e.g., by Liu et al. (2006), Buehler (2008), and Buehler and  Gao (2006). A comprehensive paper on the interface fracture mechanics is presented by Banks-Sills et al. (2015). Concerning chains in biology see, e.g., a book by Alberts et al. (2002).

The pioneering work by Thomson et al. (1971) is the first among the publications most related to the problem under consideration.  Slepyan and Troyankina (1984) studied a transition wave in a bistable chain. The dynamic splitting of a strip of two identical chains was considered analytically and numerically in Marder and Gross (1995) (the splitting under a constant load).  Mishuris et al. (2009, 2014), Ayzenberg-Stepanenko et al. (2014) and Slepyan et al. (2010) studied the splitting under a sinusoidal wave and in the presence of internal energy. The dynamical extraction of a single chain from a discrete lattice was examined in Mishuris et al. (2008). Lastly, the formulation and some results for the mode III subsonic splitting of a double-chain strip are presented in Mishuris et al. (2012).

While in the above works, {\em identical-chain systems} were considered, we now examine the quasi-static and dynamic splitting of a {\em dissimilar-chain system}. We note that similar splitting modes and phenomena can manifest themselves in both the molecular chains in biology and in macro-level composite structures. Dissimilarity changes the dynamic model dramatically. Compared with the well-studied identical-chain system (which has only a single wave speed), the considered structure represents a more complex waveguide having different physical properties. It is characterized by three wave speeds, $c_1$ and $c_2$ for the separate chains and $c_+$ for the system where the chains are connected. Accordingly, we will distinguish three wave ranges, the subsonic, $0<v<c_2<c_1$, intersonic, $c_2<v<c_+$, and supersonic, $c_+<v<c_1$. Note that the latter is subsonic for the higher speed chain. We can it the {\em supersonic splitting}, since the speed $v>c_+$ is greater than the wave speed in the intact strip.

We find that the splitting can propagate only in two of these speed ranges, the subsonic and the supersonic, whereas the steady splitting in the intersonic regime is impossible. The analytical solutions for both permitted regimes are obtained. We find the dependence of the splitting speed on the applied force and determine the quasi-static and sub- and supersonic dynamic thresholds. Also, the local-to-global speed-dependent energy release ratios defining the energy of radiated waves are presented.

Along with the chain structure in the quasi-static and dynamic regimes, we consider the corresponding continuous model. The results following from the continuum approximation can be of interest by themselves. They also present the global framework used in the analysis of the discrete system.

The quasi-static solution for the discrete model shows the role of the strip parameters (the difference in stiffness of the chains and the strength of the chain connecting links) and gives us the total energy release at low speeds (which corresponds to the energy of the rediated waves).

The steady-state solution is inadmissible if the splitting criterion is first achieved not at the splitting point, as assumed in the formulation, but ahead of it (Marder and Gross, 1995). Note that in this case, more complicated established regimes can form. The clustering and forerunning modes revealed in Mishuris et al. (2009) and  Slepyan et al. (2015), \res, are the examples.

Based on the obtained analytical results, we calculate the distribution of the chain-connecting-bond strain over the intact part of the strip, which allows us to check the admissibility.

In Conclusions, novel results presented in this paper are summarized.

\section{The strip and the basic equations}
\begin{figure}[h]
\centering
\includegraphics[scale=0.45]{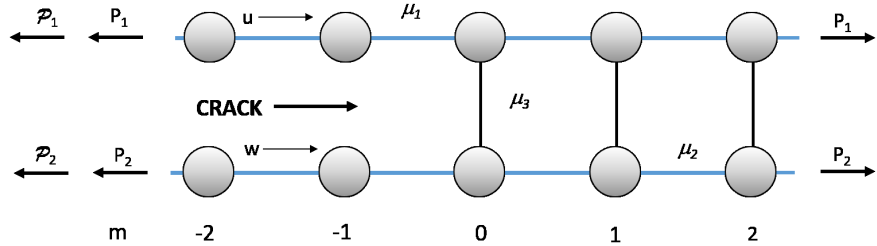}

\vspace{-3mm}
\caption{The double-chain strip under the splitting. The initial and additional forces denoted by $P_{1,2}$ and $\CP_{1,2}$, \res, are applied far from the splitting point. The strains are initially the same, so $\Gm_2P_1=\Gm_1P_2$.}
\label{f1}
\end{figure}

The strip under consideration consists of two mass-spring chains connected by elastic bonds, \fig{f1}.
One of the chains (the upper one on the Figure) is marked by number 1 as well as its parameters. Consequently, number 2 is assigned for the other chain. The connecting bonds are marked by 3. The bond stiffnesses $\Gm_i, i=1,2,3$ can be different, whereas the value of the point masses $M$ is assumed to be the same for both chains. There are three long wave speeds. The first two relate to the free chains $c_{1,2} = \sqrt{a^2\Gm_{1,2}/M}$, where $a$ is the distance between neighboring masses, and the last one, $c_+= \sqrt{a^2(\Gm_1+\Gm_2)/(2M)}$, corresponds to the strip of the connected chains. Similarly, we introduce parameter $c_3 = \sqrt{a^2\Gm_3/M}$  related to the stiffness of the chain-connecting bonds. In the following, we take $\Gm_2\le \Gm_1$ and $a$ and $M$ as the length and mass units, \res. In these terms, $\Gm_i = c_i^2$. We use notations $\Gm_i$  in the quasi-static case, and $c_i^2$ in the dynamic case. The longitudinal displacements for the upper and lower chains (as in the Figure) are denoted by $u_m(t)$ and $w_m(t)$, \res, where $m$ is the mass number. The connecting bond strain is defined as $Q_m=w_m-u_m$.

We assume that some conditions prevent transverse deviations without affecting the longitudinal displacements. In particular, it can correspond to the three-chain strip where one chain is placed between two others (each of the stiffness $\Gm_2^2/2$). Without loss of generality we set $\Gm_2\le \Gm_1$.

The homogeneous dynamic equations for the connected chains area, in terms of the wave speeds $c_i, i=1,2$, are
\beq \ddot{u}_m = c_1^2(u_{m-1}-2u_{m}+u_{m+1})-c_3^2(u_m - w_m)\,,\n
		    \ddot{w}_m = c_2^2(w_{m-1}-2w_{m}+w_{m+1})+c_3^2(u_m - w_m)\,.\eeq{1}
The separated chains obey these equations with $c_3=0$.

\section{Dynamic problem for a continuous model}\label{dpcm}
\begin{figure}[h]
\centering
\begin{subfigure}{.5\textwidth}
  \centering
  \includegraphics[width=.8\linewidth]{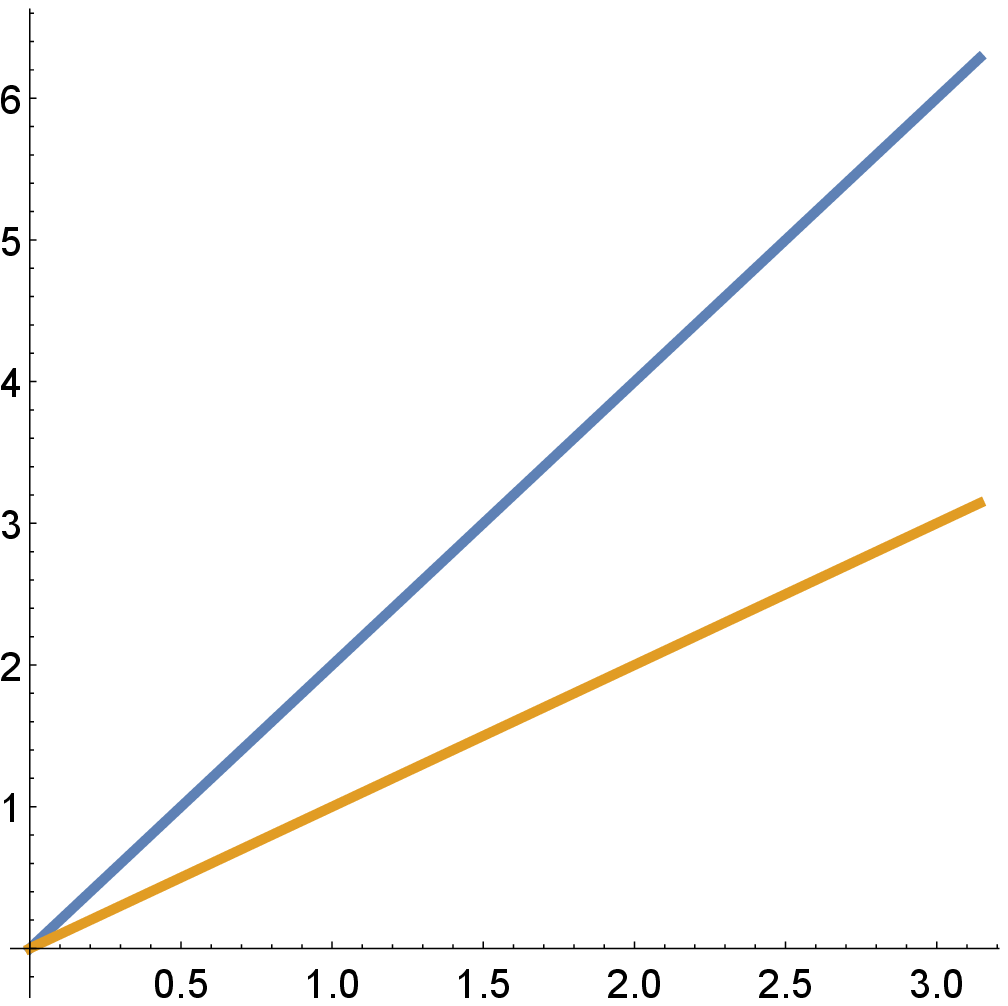}
  \vspace{4mm}
  \caption{}
  \label{2sub1}
\end{subfigure}%
\begin{subfigure}{.5\textwidth}
  \centering
  \includegraphics[width=.8\linewidth]{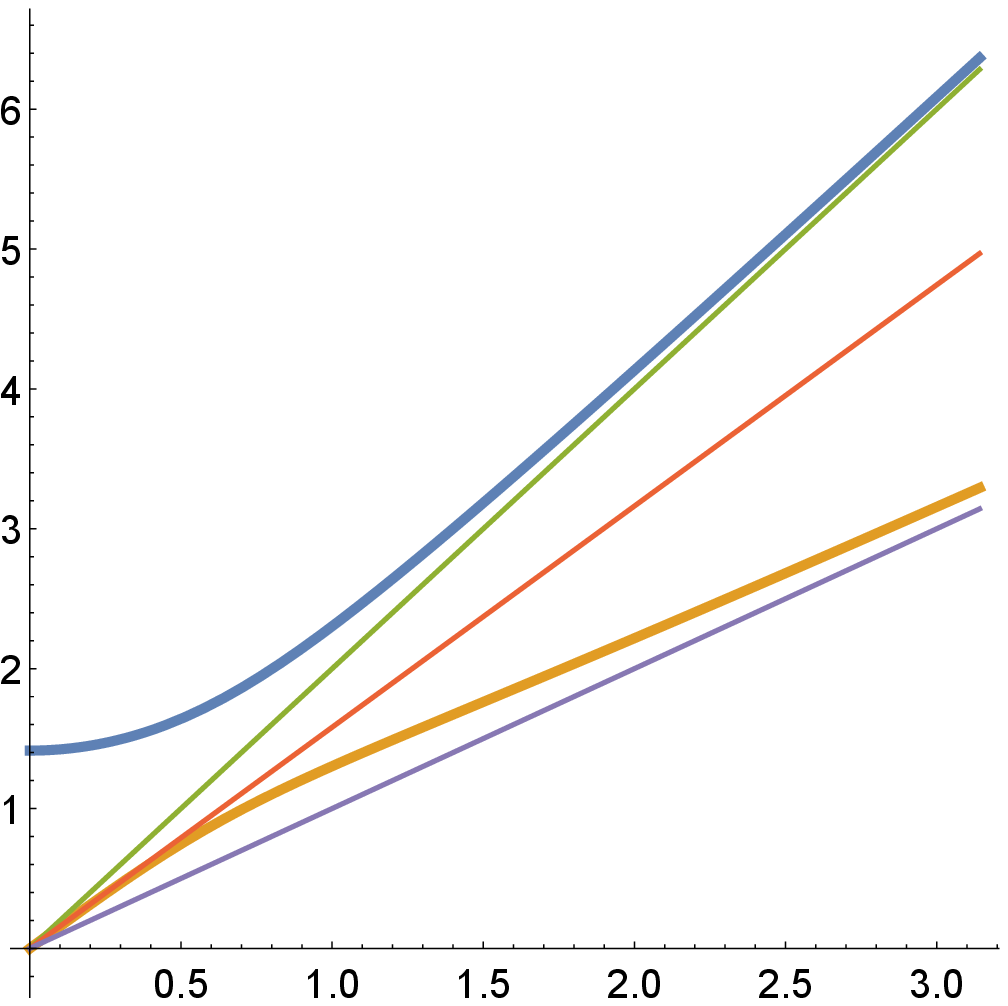}
  \vspace{4mm}
  \caption{}
  \label{2sub2}
\end{subfigure}
\begin{picture}(0,0)
	\put(-120,20){$k$}
	\put(114,20){$k$}
	\put(-45,212){$\omega_1$}
	\put(-45,130){$\omega_2$}
	\put(190,212){$\omega_1$}
	\put(190,132){$\omega_2$}
	\put(190,105){3}
	\put(190,150){2}
	\put(190,180){1}	
\end{picture}
\caption{Dispersion dependencies for the continuous model after (a) and before (b) the splitting for dissimilar strings ($c_1 = 2, c_2 = c_3 = 1$). Rays $\Go = v k$  (b) correspond to $v$ = $c_1$ (1), $c_+$ (2), $c_2$ (3). The rays are tangent to $\Go_1$  at infinity (1) $\Go_2$ at zero (2) and $\Go_2$ at infinity (3). }
\label{f2}
\end{figure}

We first consider the steady-state dynamic problem for the corresponding homogeneous elastic strings connected by the uniformly distributed elastic links. It can be considered as an independent model or as a long-wave approximation of the discrete system. This model gives us the global (macrolevel) framework for both the quasi-static and dynamic regimes of the discrete chain considered in the following sections. The equations are
\beq \ddot{u}(x,t)-c_1^2u''(x,t) +c_3^2(u(x,t)-w(x,t))=0\,,\n
\ddot{w}(x,t)-c_2^2w''(x,t) -c_3^2(u(x,t)-w(x,t))=0~~~(\Gn = x-vt >0)\eeq{0}
for the intact region and
\beq \ddot{u}(x,t)-c_1^2u''(x,t) =0\,,~~~
\ddot{w}(x,t)-c_2^2w''(x,t) ~~~(\Gn<0)\eeq{cm1}
for the region of the separated strings. The corresponding dispersion dependencies are
\beq
	\Go = \pm\Go_{1,2}\,,~~~\Go_{1,2} =\sqrt{c_+^2k^2+c_3^2 \pm \sqrt{\left(\frac{(c_1^2-c_2^2)}{2}\right)^2 k^4+c_3^4}}\,,~~~c_+= \sqrt{\f{c_1^2+c_2^2}{2}}\,,\eeq{omega4}
where $k$ and $\Go$ are the wavenumber and frequency, \res. In \fig{f2}, the dependencies are plotted in the first quadrant of the ($k,\Go$)-plane.
The dispersion relations allows us to see which waves can be emitted by the propagating splitting. The parameters of the wave, the wavenumber $k$ and frequency $\Go$, are the coordinates of the (isolated) point of intersection of the ray $\Go=kv$ with a dispersion curve. Also, the dispersion relations define the direction in which the wave propagates. Namely, the wave propagates to the right if $c_g>v $ and if the corresponding dispersion relation is valid for the area to the right, and vice versa.

In particular, it can be seen that in the considered here continuous system, no sinusoidal wave can be emitted by the moving splitting. Indeed, for any couple ($k,\Go$) with $\Go>0$ on the dispersion curves (b) the group speed $c_g < v$, and the wave must propagate to the left, but these dispersion dependencies are not valid on the left side. At the same time, no such intersection point exist on the dispersion curves (b).

\begin{figure}[h]
\centering
\begin{subfigure}{.5\textwidth}
  \centering
  \includegraphics[width=.8\linewidth]{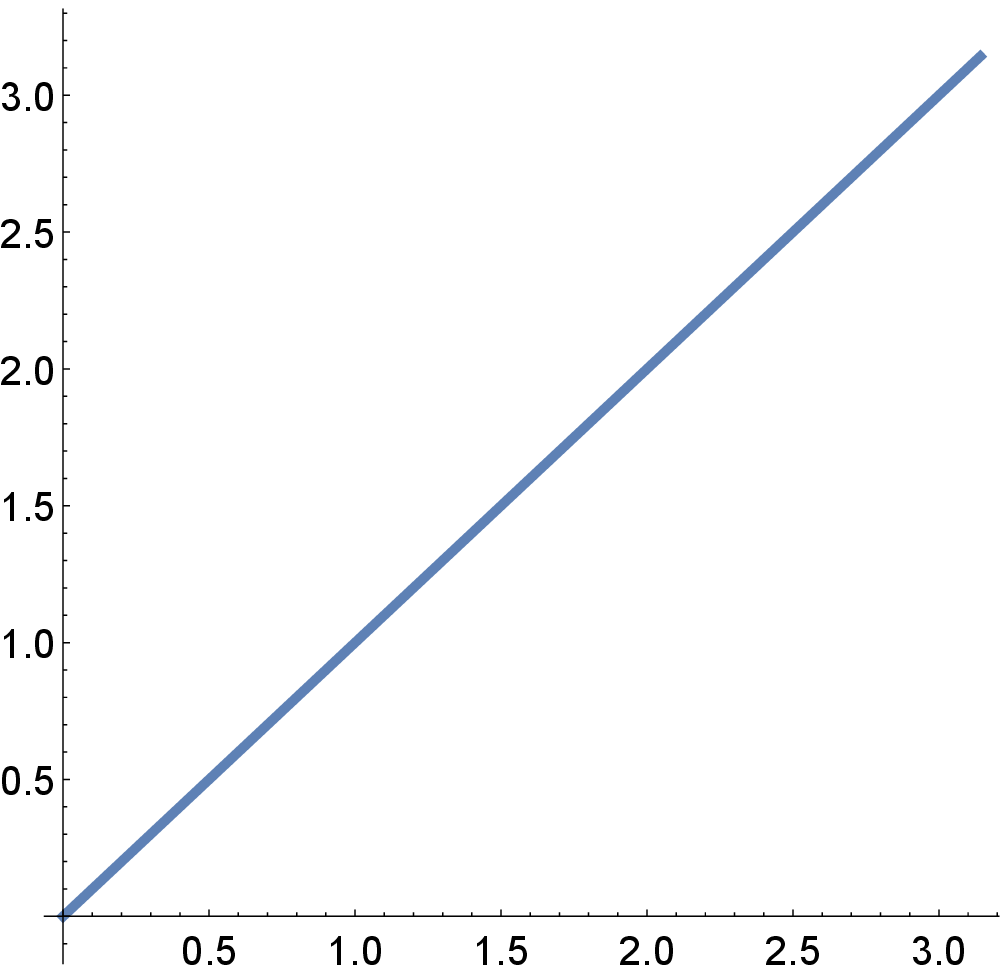}
  \vspace{4mm}
  \caption{}
  \label{3sub1}
\end{subfigure}%
\begin{subfigure}{.5\textwidth}
  \centering
  \includegraphics[width=.8\linewidth]{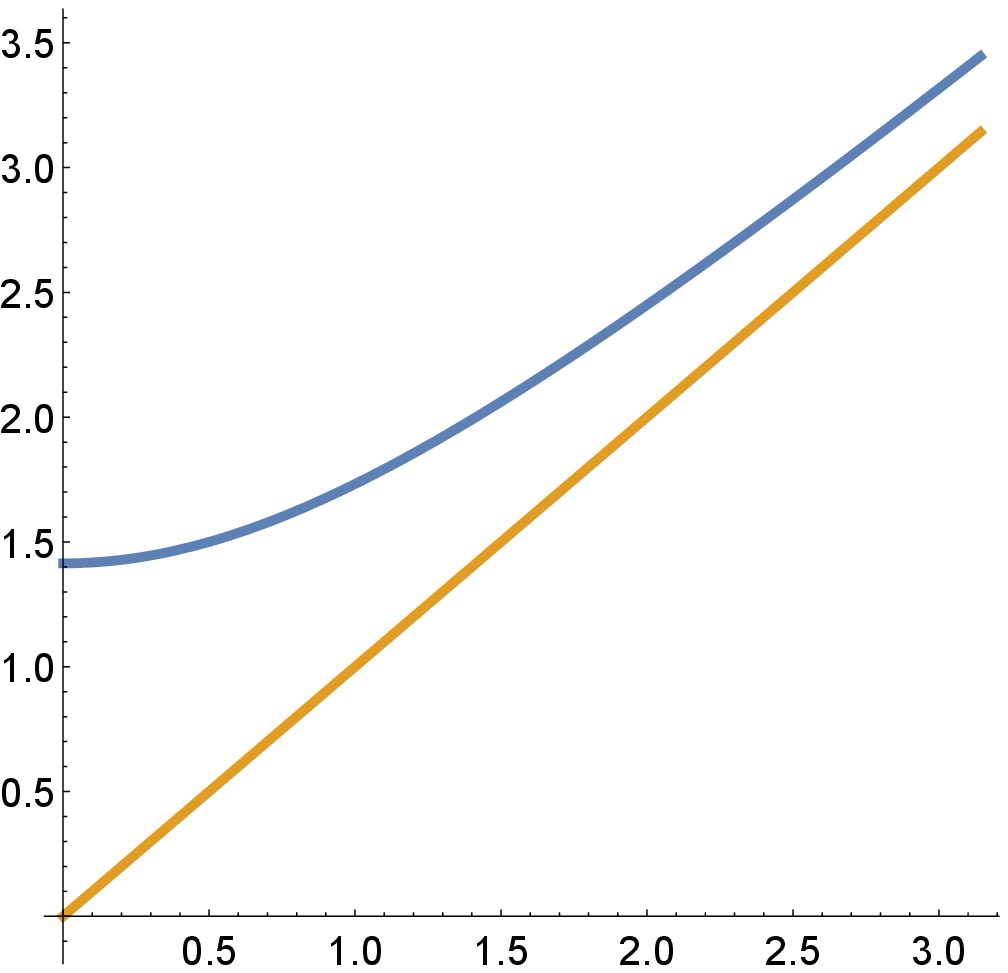}
  \vspace{4mm}
  \caption{}
  \label{3sub2}
\end{subfigure}
\begin{picture}(0,0)
	\put(-120,20){$k$}
	\put(114,20){$k$}
	\put(-45,206){$\omega_1$}
	
	\put(190,206){$\omega_1$}
	\put(190,166){$\omega_2$}
\end{picture}
\caption{Dispersion curves for the continuous model after (a) and before (b) the splitting for similar strings ($c_1 = c_2 = c_3 = 1$).}
\label{f3}
\end{figure}

Thus, in this model, the energy can be radiated from the splitting point only by waves with $k=\Go=0$. For comparison, the dispersion curves for the similar strings ($c_1=c_2$) are shown in \fig{f3}.

For the steady-state regime, the displacements are functions of the single variable $\Gn=x-vt$, and the equations \eq{0} become
\beq (c_1^2-v^2)u''(\Gn) - c_3^2(u(\Gn)-w(\Gn))=0\,,\n
(c_2^2-v^2)w''(\Gn) +c_3^2(u(\Gn)-w(\Gn))=0~~~(\Gn>0)\,;\n
u''(\Gn) = w''(\Gn) =0~~~(\Gn<0)\,.\eeq{cm2}
Among eigenvalues corresponding to the intact region there are a double zero and $\pm \nu$ with
\beq \nu = c_3\sqrt{\f{c_1^2+c_2^2-2v^2}{(c_1^2-v^2)(c_2^2-v^2)}}\,.\eeq{cm3}
The displacements can be represented as follows
\beq u(\Gn) =A_0\Gn  + B(c_2^2-v^2)\E^{-\nu\Gn} \,,~~~w(\Gn)= A_0\Gn - B(c_1^2-v^2)\E^{-\nu\Gn}~~~(\Gn>0)\,,\n
u(\Gn) = A_{10} + A_{11}\Gn\,,~~~w(\Gn) = A_{20} + A_{21}\Gn~~~(\Gn<0)\,.\eeq{cm4}
Also, we have the limiting relation for the connecting bond strain $Q(\Gn)$
\beq Q(0) = w(0)-u(0) = Q_*\,.\eeq{lsc1}
We find
\beq A_{10,20} = \pm B(c_{2,1}^2-v^2)\,,~~~A_{11} = A_0- \nu B(c_2^2-v^2)\,,~~~A_{21} = A_0+\nu B(c_1^2-v^2)\eeq{fcr1}
with
\beq B =  \f{Q_*}{2v^2 -c_1^2-c_2^2}\,.\eeq{fcr2}
These functions and their derivatives are continuous at $\Gn=0$
\beq c_1^2u'(\Gn) = \CP_1\,,~~~c_2^2w'(\Gn) = \CP_2~~~(\Gn<0)\,.\eeq{fcr3}
This equations yield the force $-$ critical strain relation
\beq \f{\CP_1}{c_1^2}-\f{\CP_2}{c_2^2} =2\GT=\nu Q_* = c_3\sqrt{\f{c_1^2+c_2^2-2v^2}{(c_1^2-v^2)(c_2^2-v^2)}}\,Q_*\eeq{fcr4}
and the value of $A_0$
\beq A_0= \f{1}{c_1^2+c_2^2-2v^2}\gl[\CP_1\gl(1-\f{v^2}{c_1^2}\gr)+\CP_2\gl(1-\f{v^2}{c_2^2}\gr)\gr]\,.\eeq{fcr5}
This expression shows that $A_0$,  as the strain at $\Gn=\infty$, is independent of the connecting bond stiffness $c_3$.

To determine the energy coming to the transition point, we consider conductive and inductive energy fluxes through the cross-sections at $\Gn<0$ and $\Gn=\infty$ (it is the same as the use of the $J$-integral reduced for our case). The energy release rates from the left and from the right are
\beq G= G_{left} + G_{right}\,,\n
G_{left} = \CP_1A_{11}+\CP_2A_{21} -\f{1}{2}[A_{11}^2(c_1^2+v^2) + A_{21}^2(c_2^2+v^2)]\,,\n
G_{right}= - A_0^2(c_1^2 +c_2^2 +2v^2)+ \f{1}{2}A_0^2(c_1^2 +c_2^2 +2v^2)\eeq{fcr6a}
It appears that this expression reduces to
\beq G= \f{2(c_1^2-v^2)(c_2^2-v^2)}{c_1^2+c_2^2-2v^2} \GT^2=\f{2c_3^2}{\nu^2}\GT^2\,.\eeq{fcr6}

\begin{figure}[h]
\centering
  \includegraphics[scale=0.8]{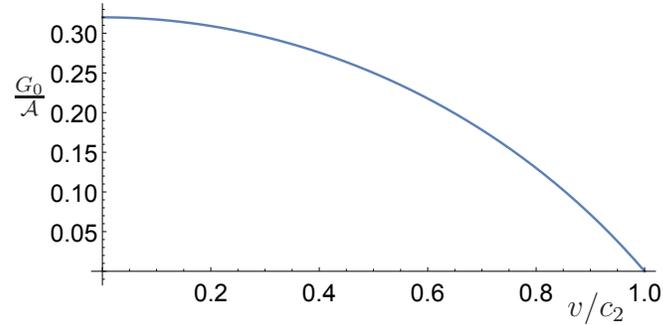}
  \vspace{4mm}
  \caption{Dynamic continuum model. Subsonic regime. The fracture energy vs. the external forces work.}
  \begin{picture}(0,0)
	\put(-130,135){$\f{G_0}{\CA}$}
	\put(80,55){$v/c_2$}		
  \end{picture}
  \label{f4}
\end{figure}

Thus, referring to \eq{fcr4} and \eq{fcr6} we find that the local and global energy releases are the same
\beq G_0=\f{1}{2}c_3^2Q_*^2 = G\,.\eeq{scr6a}
Physically, the equality $G=G_0$  follows from the fact that in this process, sinusoidal waves are not emitted. However, there are longitudinal non-oscillating waves carrying energy away from the transition point. In addition, a part of the external force work is spent for the strain energy. To determine the total radiated and strain energy rates, we compare the work of the external forces with the fracture energy. In doing so, we calculate the work on the antisymmetric strain $\GT$ taking into account that the symmetric strain does not take any part in the splitting. Referring to \eq{fcr4}, the work is
\beq \CA= (c_1^2+c_2^2)\GT^2 = \Ga(v)G_0\,,~~~\Ga(v)=\f{(c_1^2+c_2^2)(c_1^2+c_2^2-2v^2)}{2(c_1^2-v^2)(c_2^2-v^2)}\,.\eeq{w1a}
The ratio $G_0/\CA$  vs. speed $v$ in the subsonic speed range, $0\le v < c_2$ is shown in \fig{f4}

\subsection{Supersonic regimes}\label{ir}
 As can be seen in \eq{cm3}, $\nu$ is imaginary in the intersonic range and real in the supersonic one, which corresponds to sinusoidal and exponential waves, \res. It follows that in the former, the connecting bond strain $Q(\Gn)$  does not decrease with the distance from the splitting point, and the value $Q(0)$ is periodically repeated or even exceeded as $\Gn$ grows. Such a solution does not satisfy the admissibility condition; it contradicts the assumption that the bond breaks at $\Gn=0$ but not at $\Gn>0$. Thus, the steady-state regime in the intersonic speed range is forbidden.

In the expressions \eq{cm4} for the supersonic speed range $c_+<v<c_1$, we have to put $A_0=0$, since the waves induced at the left vanish as $\Gn \to \infty$. Also, the second relation in \eq{fcr3} does not hold since $c_2<v$. We have
\beq u(\Gn) =-B(v^2 - c_2^2)\E^{-\nu\Gn} \,,~~~w(\Gn)= -  B(c_1^2-v^2)\E^{-\nu\Gn}~~~(\Gn>0)\,,\n
u(\Gn) =A_{1,0}+ A_{11}\Gn\,,~~~w(\Gn) = A_{20} + A_{21}\Gn~~~(\Gn<0)\eeq{cm4is}
with
\beq A_{10} = - B(v^2-c_{2}^2)\,,~~A_{20} = - B(c_{1}^2-v^2)\,,~~A_{11} = \nu B(v^2 -c_2^2)\,,~~A_{21} = \nu B(c_1^2-v^2)\,,\n
 B =\f{Q_*}{2v^2-c_1^2-c_2^2}\,,~~~2v^2>c_1^2+c_2^2~~\Longrightarrow~~c_1^2-v^2<v^2-c_2^2\,.~~~\eeq{fcr1iz}
The tensile forces at the left are
\beq \CP_1=c_1^2\nu B(v^2-c_2^2)=\f{c_3c_1^2Q_*}{\sqrt{2v^2-c_1^2-c_2^2}}\sqrt{\f{v^2-c_2^2}{c_1^2-v^2}}\,,\n
\mathbb{P}_2= c_2^2 \nu B(c_1^2-v^2) =\f{c_2^2 (c_1^2-v^2)}{c_1^2(v^2-c_2^2)}<\CP_1~~~(\Gn<0)\,.\eeq{p1o}

Note that the value of $\CP_1$ corresponds to the external force acting on the first string, while $\mathbb{P}_2$ is the tensile force appeared in the second string due to the action of the first string. As can be seen in \eq{cm4is} and \eq{fcr1iz}, both strains, $A_{11}$ and $A_{21}$ are positive (if $\CP_1 >0$), although $A_{21}$  strain is caused by the action of the left-directed force. This fact, however, is typical for the supersonic speed of the load, see, e.g., Langlet et al. (2012). Recall that any external force applied to second string far away from the splitting point cannot be detected in the considered area since $c_2<v$.

A U-shaped curve corresponds to the function $\CP_1(v)$, and its minimum is defined by the relation
\beq  \f{\CP_{1,min}}{c_3^2Q_*} = \f{(1+\sqrt{2})c_{12}^2}{\sqrt{c_{12}^2 -1}}\,,\eeq{cpm1}
where $c_{12}=c_1/c_2$. It is achieved at
\beq \f{v}{c_2} = \sqrt{1 +\f{\sqrt{2}}{2}(c_{12}^2-1)}\,.\eeq{cpm2}

\begin{figure}[h]
\centering
\begin{subfigure}{.5\textwidth}
  \centering
  \includegraphics[width=.83\linewidth]{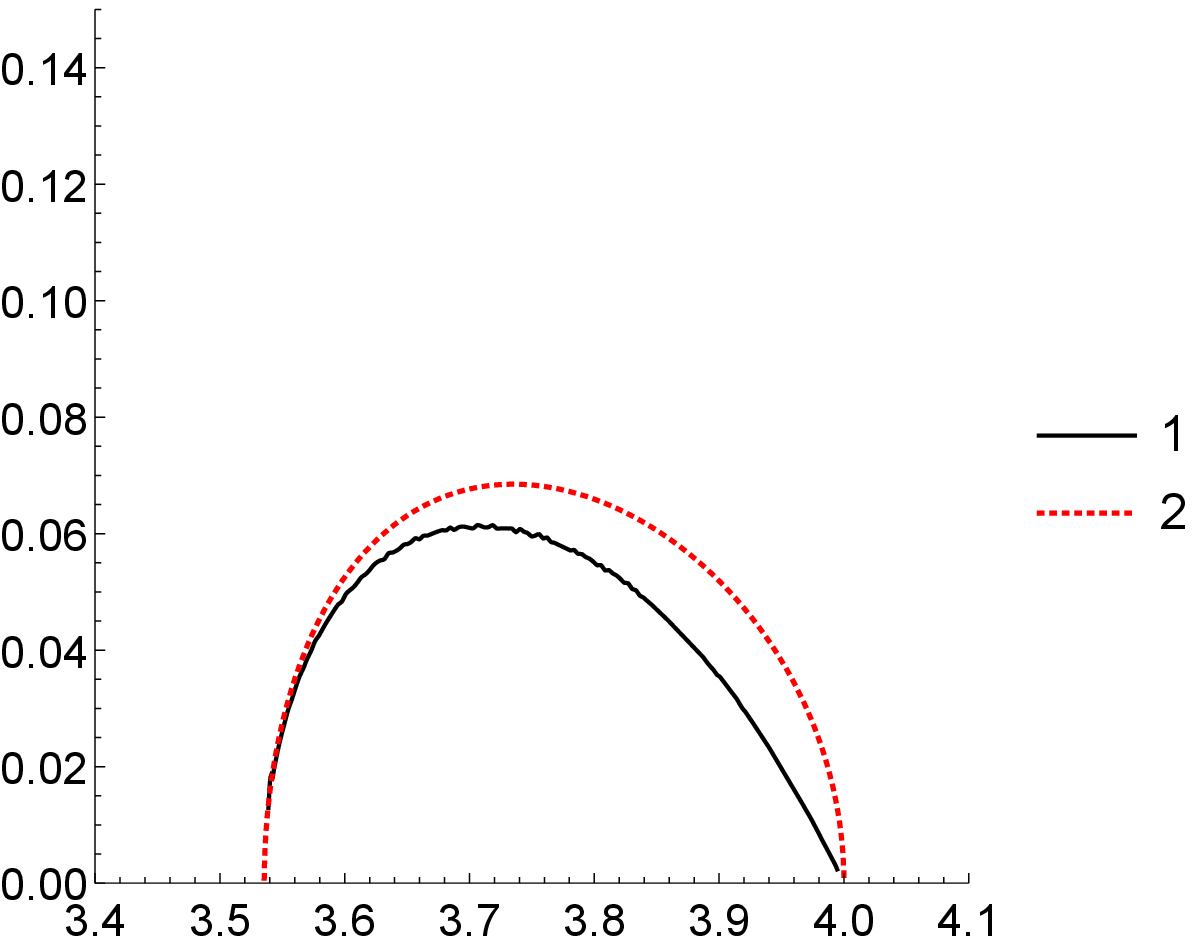}
  \vspace{4mm}
  \caption{$c_1 = 4,\qquad c_2 = 3,\qquad c_3 = 1$}
  \label{10sub1}
\end{subfigure}%
\begin{subfigure}{.5\textwidth}
  \centering
  \includegraphics[width=.8\linewidth]{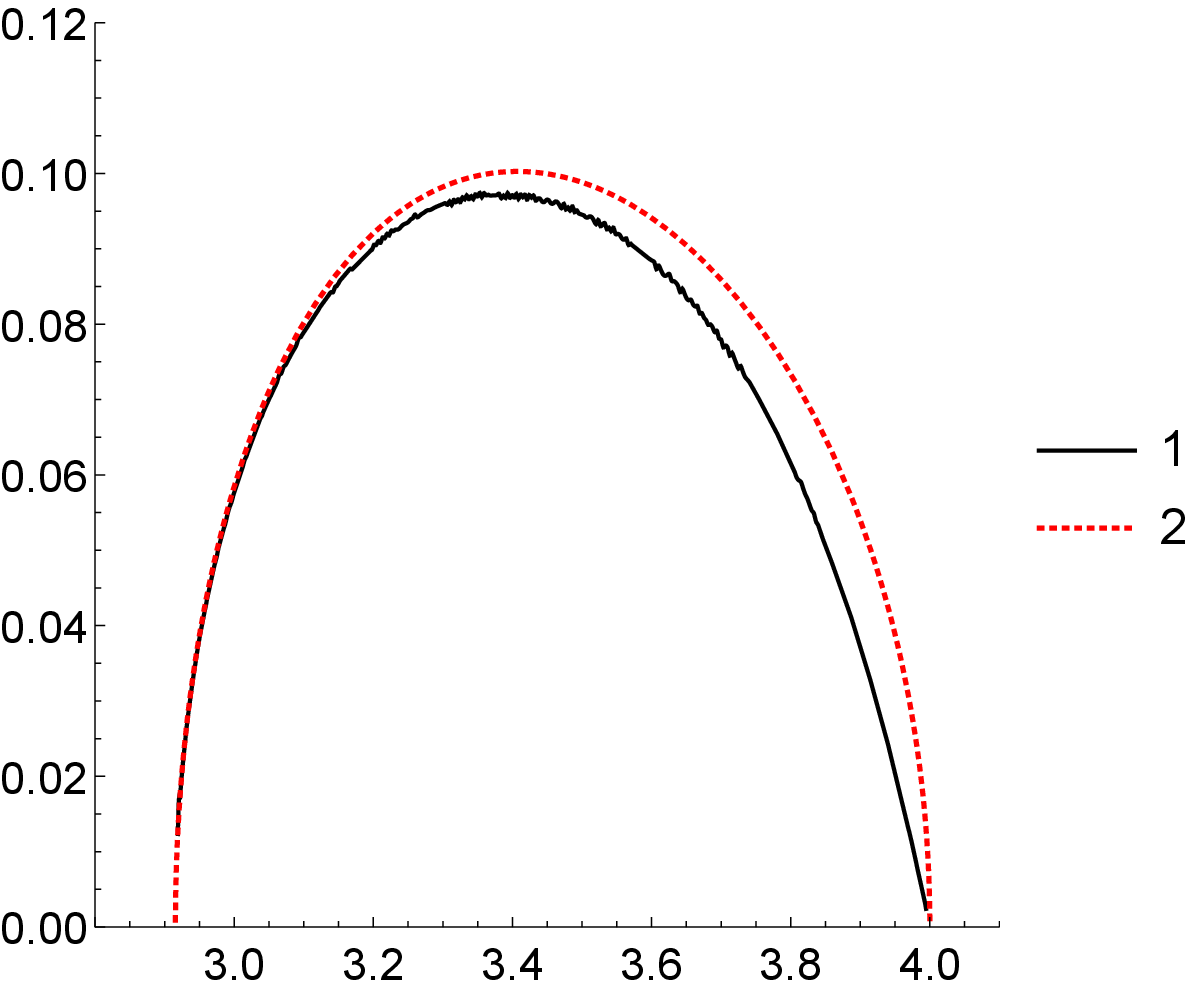}
  \vspace{4mm}
  \caption{$c_1 = 4,\qquad c_2 = 1,\qquad c_3 = 1$}
  \label{10sub2}
\end{subfigure}
\begin{picture}(0,0)
	\put(-233,143){$\f{c_3^2 Q_*}{ \CP_1}$}
	\put( 6,143){$\f{c_3^2 Q_*}{\CP_1}$}
	\put(-120,20){$v$}
	\put(114,20){$v$}
\end{picture}
\caption{Supersonic regime. Dynamic discrete model (1),  dynamic continuous model (2).}
\label{f5}
\end{figure}

The relation between the external force and the critical tensile force in the connecting bond $c_3^2Q_*$  is shown in \fig{f5}, where the corresponding results following from the continuous and discrete dynamic formulations are presented too.

It is remarkable that in the supersonic regime, contrary to the subsonic one, the discreteness does not play a significant role. It also deserves to be noted that the dynamic threshold, as the minimum of $\CP_1$, for the supersonic mode becomes less than the corresponding static value for $c_1/c_2>1.518653687 $  (see \az{tdt11}).

\subsubsection{The thresholds}\label{tdt11}
\begin{figure}[h]
\centering
  \includegraphics[scale=0.4]{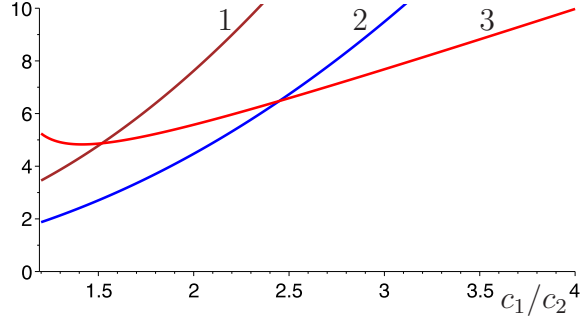}
  \vspace{4mm}
  \begin{picture}(0,0)
\put(-36,0){$c_1/c_2$}		
\put(-143,105){$1$}		
\put(-92,105){$2$}		
\put(-44,105){$3$}		
  \end{picture}
  \caption{The splitting thresholds as minimal ratios $\CP_1/(c_3^2Q_*)$  with $\CP_2=0$ for the discrete model in statics (1) and the continuous model in dynamics at the subsonic (2) and supersonic (3) speed ranges ($c_{32}=1$).}
  \label{f6}
\end{figure}

\begin{figure}[h]
\centering
  \includegraphics[scale=0.4]{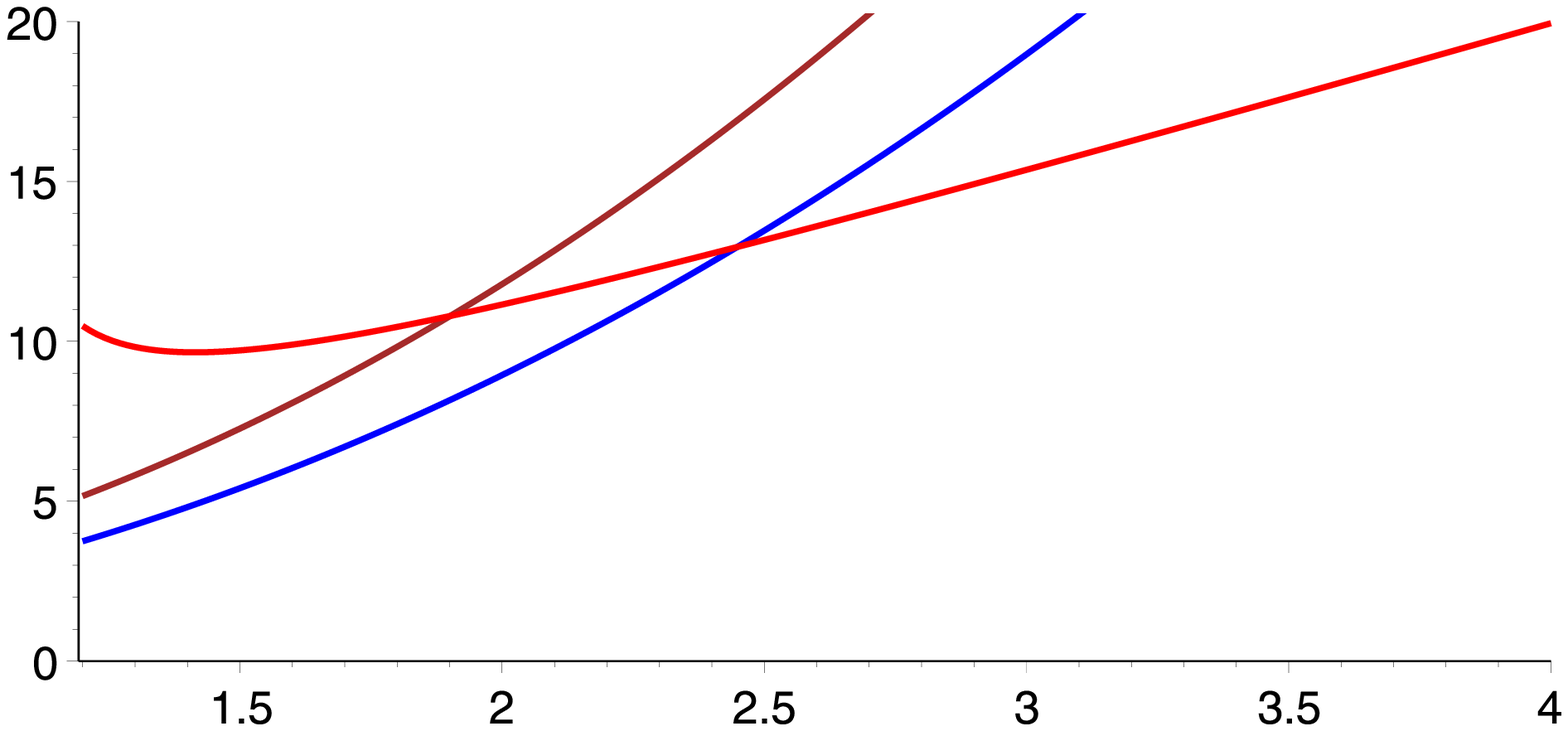}
  \vspace{4mm}
  \begin{picture}(0,0)
\put(-37,-2){$c_1/c_2$}		
\put(-117,97){$1$}		
\put(-88,97){$2$}		
\put(-34,97){$3$}		
  \end{picture}
  \caption{The splitting thresholds as minimal ratios $\CP_1/(c_3^2Q_*)$ with $\CP_2=0$ for the discrete model in statics (1) and the continuous model in dynamics at the subsonic (2) and supersonic (3) speed ranges ($c_{32}=1/2$).}
  \label{f7}
\end{figure}

\begin{figure}[h]
\centering
  \includegraphics[scale=0.8]{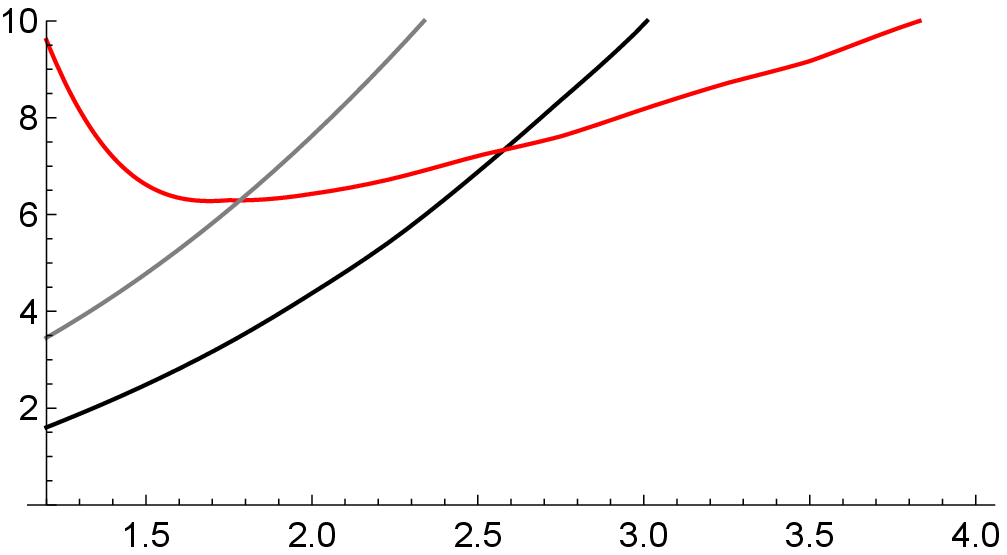}
  \vspace{4mm}
  \begin{picture}(0,0)
\put(-41,-2){$c_1/c_2$}		
\put(-159,107){$1$}		
\put(-111,107){$2$}		
\put(-82,107){$3$}		
  \end{picture}
  \caption{The splitting thresholds as minimal ratios $\CP_1/(c_3^2Q_*)$ with $\CP_2=0$ for the discrete model in statics (1) and dynamics at the subsonic (2) and supersonic (3) speed ranges ($c_{32}=1$).}
  \label{f7d}
\end{figure}

We now compare the thresholds corresponding to the quasi-static, subsonic and supersonic regimes \eq{cpm1}. To consider all cases under the same conditions, we take $\CP_2=0$. As follows from \eq{fcr4}, the subsonic value
\beq \f{\CP_{1}^{ss}}{c_3^2Q_*} = \f{c_{12}}{c_{32}}\sqrt{1+c_{12}^2}\eeq{cpm1sub}
reaches at $v=0$, whereas the discrete quasi-static value follows from \eq{tis62} with \eq{tis1} (see the next Section) as
\beq \f{\CP_{1}^{qs}}{c_3^2Q_*} = (\Gl_3-1)\f{c_{12}^2}{c_{32}}\,.\eeq{cp1st}
These three dependencies of the critical force $\CP_1$  are plotted in \fig{f6} and \fig{f7} for $c_{32}=1$ and $c_{32}=1/2$, \res.  The supersonic threshold is below the discrete model static value for $c_1/c_2> 1.518653687 $ and $c_1/c_2>1.900953739$  for these cases, \res. It is below the continuous model subsonic value for $c_1/c_2> 2.448515161$ for both cases. Similar dependencies obtained in \az{tdpds} for the discrete system are presented in \fig{f7d} for  $c_{32}=1$. The critical ratios (in the above sense) are $c_1/c_2= 1.783585043$  and  $c_1/c_2= 2.578581133$, \res.

\section{The quasi-static problem for the discrete strip}\label{tsp}
\subsection{General relations}
The equation \eq{1} admits the following static solution
\beq u_{m+1} =\Gl u_m\,,~~~w_{m+1} =\Gl w_m\,,~~~\Gl=\Gl_{1,2}=1\,,\n
\Gl=\Gl_{3,4}=1+\Gf \pm\sqrt{\Gf^2 +2\Gf}\,,~~~\Gl_4=\f{1}{\Gl_3}<1\,, ~~~\Gf=\f{\Gm_3(\Gm_1+\Gm_2)}{2\Gm_1\Gm_2}\eeq{tis1}
with
\beq w_m=u_m ~~~(\Gl=\Gl_{1,2})\,,~~~\Gm_2w_m = - \Gm_1u_m~~~(\Gl = \Gl_{3,4})\,.\eeq{tis3}
Note that the former of the above equalities reflects uniform strain equal for both chains, whereas the latter corresponds to the tensile forces different by sign. The general expressions for the displacements of the intact strip are
\beq u_m = C_1 + C_2 m + C_3\Gm_2\Gl_3^m + C_4\Gm_2\Gl_4^m\,,\n
w_m = C_1 +C_2 m -C_3\Gm_1\Gl_3^m -C_4\Gm_1\Gl_4^m\,,\eeq{tis4}
where $C_{1,...,4}$ are arbitrary constants, and the constant $C_2$ represents the uniform strain.
The $C_{3,4}-$terms represent the boundary effects for a finite or semi-infinite chain. For the infinite intact strip ($m=0, \pm 1, ...$) we assume, as usual, that the displacements do not grow exponentially as $m\to\pm\infty$; hence the constants $C_{3,4}=0$ and the initial tensile forces are
\beq P_1= C_2 \Gm_1\,,~~~P_2 =C_2 \Gm_2\,, u_m=w_m\eeq{tis5}
with the $\Gm_3$-bonds unstressed.

\subsection{The splitting problem}
Consider the equilibrium of the unbounded chains connected at $m\ge 0$ and separated at $m<0$. In the latter region, the initial uniform tensile forces \eq{tis5} can exist together with arbitrary additional forces $\CP_{1,2}$. For the static case, we assume that the equilibrium is supported by the additional remote forces (with the principle value $-\CP_1-\CP_2$) applied to the strip at the right. The task is to determine the maximal strain in the connecting bonds. In the general solution \eq{tis4}, $C_1$ as a rigid displacement can be ignored as well as the exponentially growing $C_3-$term. So, the connecting bond strains are
\beq Q_m= w_m-u_m = Q_0\Gl_4^m\,,~~~Q_0= - C_4(\Gm_1+\Gm_2)\,.\eeq{tis57}
The displacements can be represented as
\beq  u_m = C_2m - \f{\Gm_2 Q_0}{\Gm_1+\Gm_2}\Gl_4^m\,,~~~w_m = C_2m +\f{\Gm_1 Q_0}{\Gm_1+\Gm_2}\Gl_4^m~~~(m\ge 0)\,,\n
u_m = - \f{\Gm_2 Q_0}{\Gm_1+\Gm_2} +\f{\CP_1+P_1}{\Gm_1}m\,,~~~w_m = \f{\Gm_2 Q_0}{\Gm_1+\Gm_2} +\f{\CP_2+P_2}{\Gm_2}m~~~(m\le 0)\,.\eeq{tis58}
The equilibrium equation for the points at $m=0$ are
\beq \CP_1+P_1=\Gm_3 Q_0 +(1-\Gl_4)\f{\Gm_1\Gm_2}{\Gm_1+\Gm_2}Q_0+\Gm_1C_2\,,\n
\CP_2+P_2=-\Gm_3 Q_0 -(1-\Gl_4)\f{\Gm_1\Gm_2}{\Gm_1+\Gm_2}Q_0+\Gm_2C_2\eeq{tis60}
with
\beq \Gm_2P_1=\Gm_1P_2\,.\eeq{tis600}
We find
\beq C_2=\f{P_1}{\Gm_1}+\f{\CP_1+\CP_2}{\Gm_1+\Gm_2}\eeq{tis601}
and
\beq \GT =\f{1}{2}\gl( \f{\CP_1}{\Gm_1}-\f{\CP_2}{\Gm_2}\gr) = \f{1}{2}(\Gl_3-1)Q_0\,.\eeq{tis602}

\subsubsection{Critical forces and the energy}
Thus, the expression for the critical force is
\beq \GT =\f{1}{2}(\Gl_3-1)Q_*\,,\eeq{tis62}
where $Q_*$ is the limiting strain of the chain-connecting bonds. The strain energy of the latter at breakage is
\beq G_0 = \f{\Gm_3 Q_*^2}{2}=\f{\Gm_3(1-\Gl_4)^2}{2\Gf^2}\GT^2\,.\eeq{qs5}
The global energy release rate is equal to that obtained for the continuum model \eq{fcr6} with $v=0$
\beq G = \f{2\Gm_1\Gm_2}{\Gm_1+\Gm_2}\GT^2\,,\eeq{qs6}
and the energy release ratio is
\beq \f{G_0}{G}=\Gl_4\,.\eeq{qsa6}
Referring to \eq{tis1}, this ratio is a monotonically decreasing function of $\Gf$ and
\beq \f{G_0}{G}  \to 1 ~~(\Gf\to 0)\,,~~~ \f{G_0}{G} \to 0~~(\Gf\to\infty)\,.\eeq{err1}
For $\Gf>0$ a part of the global energy release rate is spent the fracture energy itself ($G_0$), while the other part disappears. The quasi-static formulation gives no answer where does the latter go. The dynamic formulation used below shows that this is the energy of waves emitted by a slowly propagating splitting.

\section{The dynamic problem for the discrete strip}\label{tdpds}
\subsection{General relations}
\begin{figure}[h]
\centering
\begin{subfigure}{.5\textwidth}
  \centering
  \includegraphics[width=.8\linewidth]{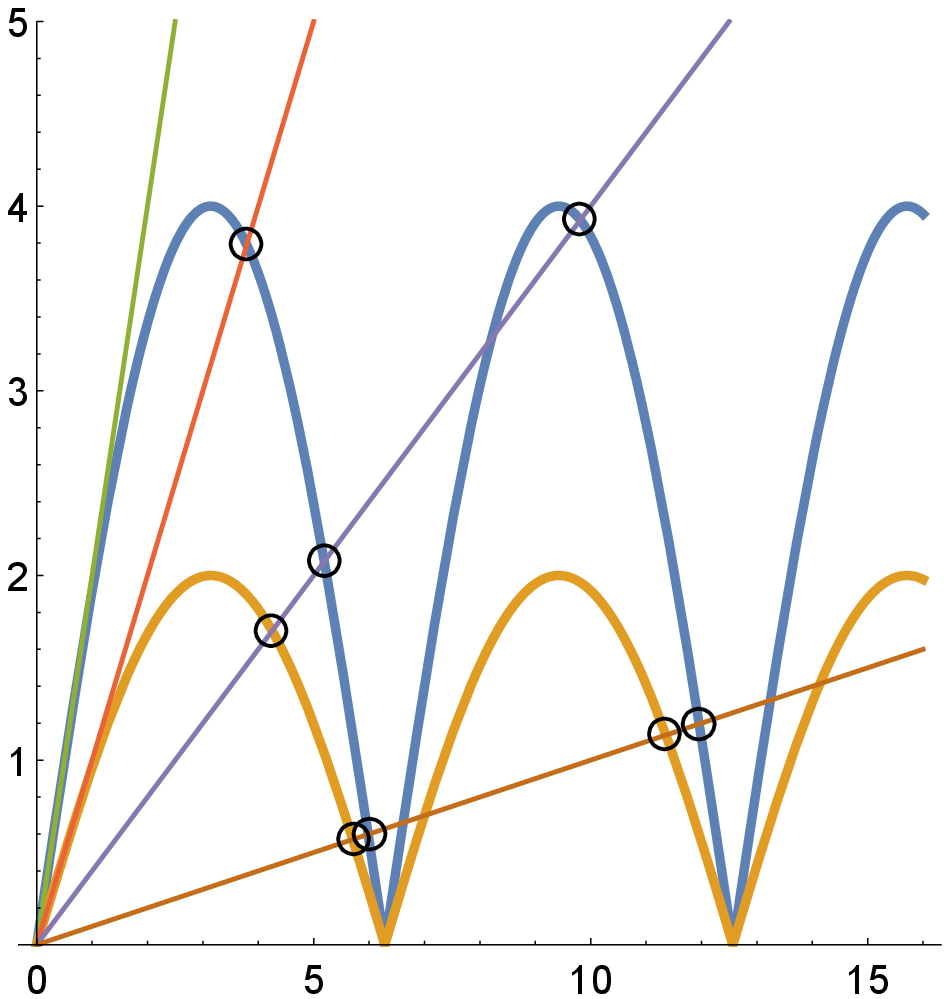}
  \vspace{4mm}
  \caption{}
  \label{5sub1}
\end{subfigure}%
\begin{subfigure}{.5\textwidth}
  \centering
  \includegraphics[width=.8\linewidth]{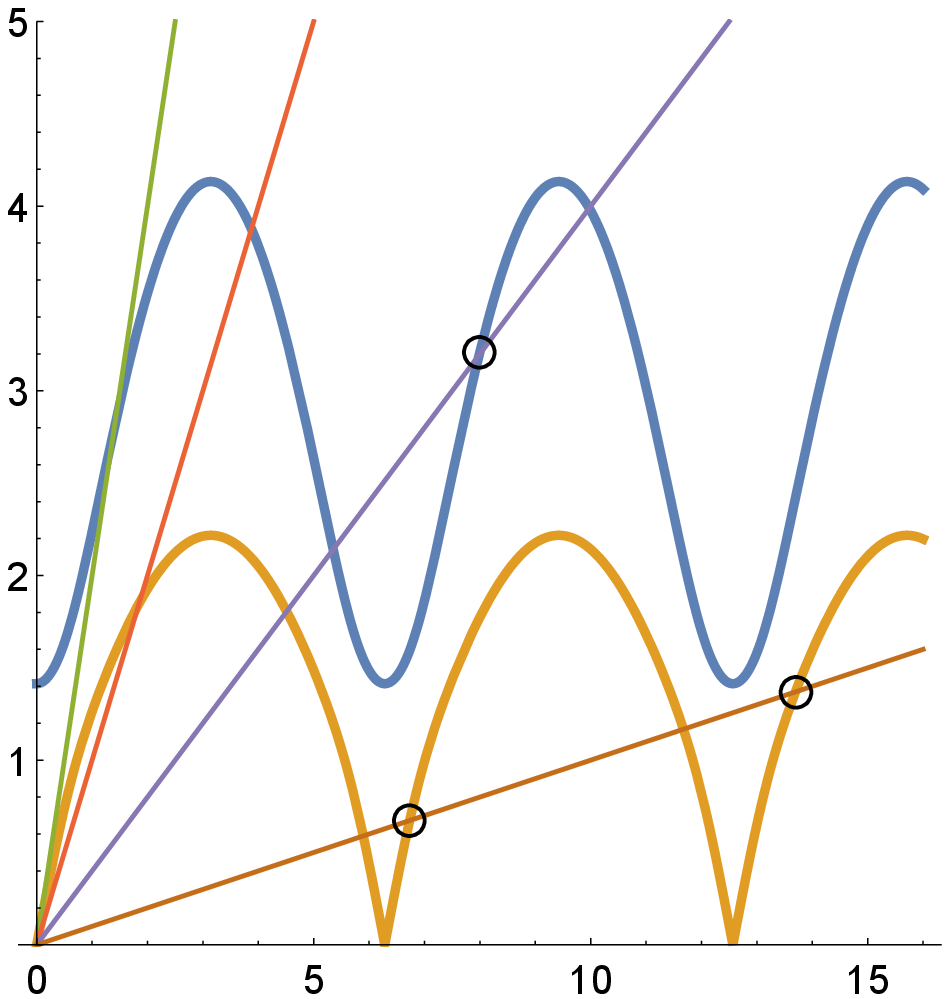}
  \vspace{4mm}
  \caption{}
  \label{5sub2}
\end{subfigure}
\begin{picture}(0,0)
	\put(-120,20){$k$}
	\put(114,20){$k$}
	\put(-45,196){$\omega_1$}
	\put(-45,121){$\omega_2$}
	\put(190,200){$\omega_1$}
	\put(190,129){$\omega_2$}
	\put(-37,103){4}
	\put(-92,205){3}
	\put(-165,205){2}
	\put(-190,205){1}	
	\put(190,100){4}
	\put(142,205){3}
	\put(69,205){2}
	\put(45,205){1}	
\end{picture}
\caption{Dispersion curves for the discrete dynamic model after (a) and before (b) the splitting for the dissimilar chains, $c_1 = 2, c_2 = c_3 = 1$. Rays $\Go = v k$ correspond to $v = 2$ (1), $1$ (2), $0.4$ (3) and $0.1$ (4). The points marked by small circles correspond to ($k,\Go$)-parameters of the sinusoidal waves radiated by the propagating splitting. The waves propagate to the left if group speed $\D \Go/\D k< v$ (marked in (a)), and vice versa (b).}
\label{f8}
\end{figure}

\begin{figure}[h]
\centering
\begin{subfigure}{.5\textwidth}
  \centering
  \includegraphics[width=.8\linewidth]{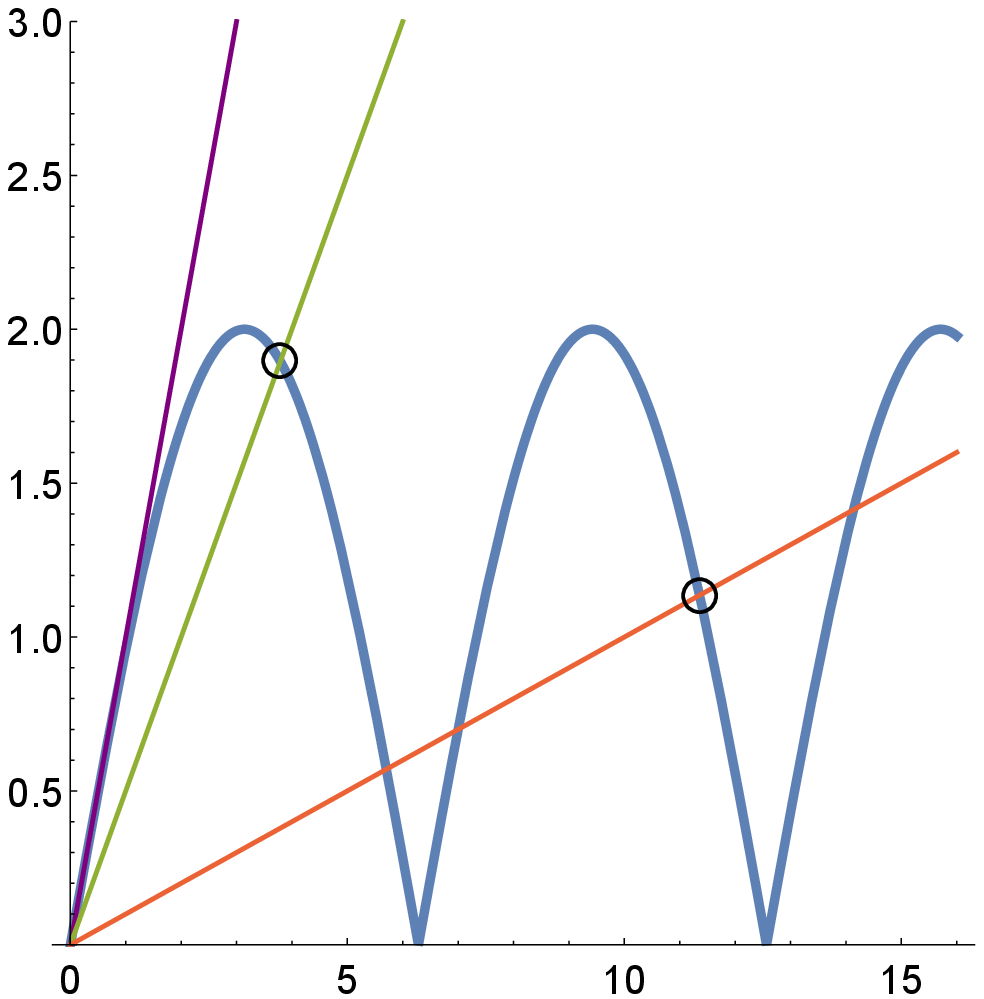}
  \vspace{4mm}
  \caption{}
  \label{6sub1}
\end{subfigure}%
\begin{subfigure}{.5\textwidth}
  \centering
  \includegraphics[width=.8\linewidth]{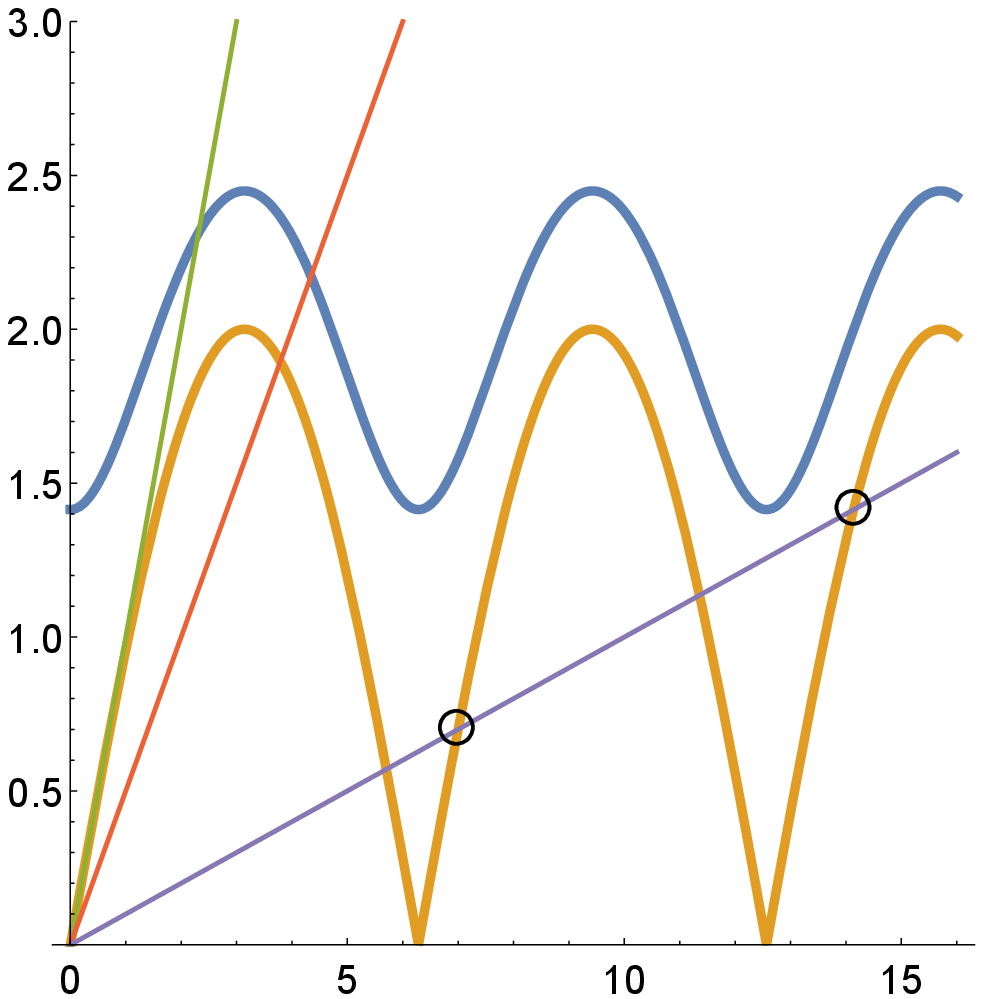}
  \vspace{4mm}
  \caption{}
  \label{6sub2}
\end{subfigure}
\begin{picture}(0,0)
	\put(-120,20){$k$}
	\put(114,20){$k$}
	\put(-45,165){$\omega_1$}
	\put(195,192){$\omega_1$}
	\put(195,165){$\omega_2$}
	\put(-37,138){3}
	\put(-150,205){2}
	\put(-185,205){1}	
	\put(195,137){3}
	\put(85,205){2}
	\put(55,205){1}	
\end{picture}
\caption{Dispersion curves for the discrete model after (a) and before (b) the splitting for identical chains ($c_1 = c_2 = c_3 = 1$). Rays $\omega = v k$ correspond to $v = 1$ (1), $0.5$ (2) and $0.1$ (3).}
\label{f9}
\end{figure}

We now consider the same strip as in \az{tsp} but with inertia of the masses (or the same as in \az{dpcm} but for the corresponding discrete structure). The dispersion relations corresponding to the equations \eq{1} are shown in \fig{f8} and \fig{f9}.

It can be seen that in the case of the discrete strip, sinusoidal waves exist in the whole range of the splitting speeds. It follows that the global energy release always exceeds the local one.

For the steady-state problem where the displacements depends only on the single variable $\Gn =m-vt$, equations \eq{1} become
\beq v^2u''(\Gn)
= c_1^2[u(\Gn-1)-2u(\Gn)+u(\Gn+1)] - c_3^2[u(\Gn) - w(\Gn)]\,,  \n
v^2w''(\Gn)
= c_2^2[w(\Gn-1)-2w(\Gn)+w(\Gn+1)]+c_3^2[u(\Gn) - w(\Gn)]\,.\eeq{ss1}
In particular, equation \eq{ss1} admits the uniform strain
\beq u(\Gn) = w(\Gn)= C_1+C_2 \Gn\,.\eeq{ss2}
The boundary-type part of the total solution is determined below.

\subsection{The Wiener-Hopf equation}
 The Fourier transform of \eq{ss1} for the intact structure leads to
\beq h_1u^F(k) - c_3^2Q^F(k)=0\,,~~~h_2 w^F(k) + c_3^2Q^F(k)=0\,,\n
 Q(\Gn)=w(\Gn)-u(\Gn)\,,~~~h_{1,2}=(s+\I k v)^2 +2c_{1,2}^2(1-\cos k)\,.\eeq{wh1}
Note that we use the so-called generalised Fourier transform. Namely, the transform of a function $f(\Gn)$ is
\beq f^F(k) = f_+(0-\I k) + f_-(0+\I k)\,,\n f_+ (s- \I k)= \int_0^\infty f(\Gn)\E^{-(s-\I k)}\D \Gn\,,~~~
f_- (s+ \I k)= \int_{-\infty}^0 f(\Gn)\E^{(s+\I k)}\D \Gn~~(0<s\to 0)\,.\eeq{gft1}

For the region where the connecting bonds are broken, $\Gn<0$, we have to introduce compensation forces $\pm c_3^2 Q_-(k)$ acting on the first and the second chains, \res. We have
\beq h_1u^F(k) - c_3^2Q_+(k) =0\,,~~~ h_2w^F(k) + c_3^2Q_+(k) = 0\,,\eeq{wh2}
and
\beq  w^F(k)-u^F(k)=Q_++Q_- = - c_3^2Q_+\gl(\f{1}{h_1}+\f{1}{h_2}\gr)\,.\eeq{wh4}
Thus, the homogeneous Wiener-Hopf equation can be represented as
\beq L(k)Q_+ +Q_- =0\eeq{wh5}
with the kernel
\beq L(k) =\f{h_1h_2 +c_3^2(h_1+h_2)}{h_1h_2}\,.\eeq{wh6}

We first consider the subsonic case: $0<v<c_2$.
Using the asymptotes ($k\to 0, s\to 0$)
\beq h_{1,2}\sim (s-(c_{1,2}-v)\I k)(s+(c_{1,2}+v)\I k)~~~(0<v<c_{1,2})\,,\eeq{wh7}
we normalise the kernel as follows:
\beq L(k) = S(k)L^0(k)\,\n
S(k)=\f{k^2+\nu^2}{(0+\I k)(0-\I k)}\,,\eeq{wh8}
where $\nu$ is defined in \eq{cm3}. We have
\beq L^0(0)=L^0(\pm \infty)=1\,,~~~ \mbox{ Ind} L^0 =0\,.\eeq{wh8a}

\subsection{Solution}
We factorise $L^0(k)$ using the Cauchy-type integral
\beq L^0_\pm (k) = \exp\gl[\pm\f{1}{2\pi\I}\int_{-\infty}^\infty \f{\mbox{Ln} L^0(\xi)}{\xi -k}\D\xi\gr]~~~(\pm\Im k >0)\,,\n
L_\pm^0(\pm\I\infty)=1\,,~~~L^0_\pm(0) =\exp\gl[\pm\f{1}{\pi}\int_0^\infty \f{\mbox{Arg} L(\xi)}{\xi}\D\xi\gr], \eeq{wh9}
and represent the Wiener-Hopf equation \eq{wh5} in the form
\beq
\f{\nu-\I k}{0-\I k}L^0_+(k)Q_+(k) +\f{0+\I k}{(\nu+\I k)L^0_-(k)}Q_-(k) =0\,.\eeq{wh10h}

In addition to this equation, there are conditions for non-oscillation part of $Q(\Gn)$ for $\Gn\to\pm\infty$
\beq Q(\infty)=0\,,~~~Q'(-\infty) = 2\GT \equiv \f{\CP_1}{c_1^2} -   \f{\CP_2}{c_2^2}\,.\eeq{bc1}
To satisfy these conditions we introduce, as usual, the delta-function (in its analytical representation) in the right-hand side of the equation \eq{wh10h}. We rewrite it in the form
\beq \f{\nu-\I k}{0-\I k}L^0_+(k)Q_+(k) +\f{0+\I k}{(\nu+\I k)L^0_-(k)}Q_-(k) =C 2\pi \Gd(k) =\f{C}{0+\I k} + \f{C}{0-\I k}\,.\eeq{wh10}
It follows that
\beq Q_+(k) =\f{C}{(\nu-\I k)L^0_+(k)}\,,~~~Q_-(k) =\f{(\nu-\I k) L^0_+(k)C}{(0+\I k)^2}\,.\eeq{wh10aa}

The constant $C$ can be determined as follows. From the condition in \eq{bc1} for $Q'(-\infty)$ we find that
\beq Q_- \sim \f{2\GT}{(0+\I k)^2}~~~(k\to 0)\,.\eeq{wh110}
Substituting this in \eq{wh10}
\beq C =  \nu L^0_+(0)Q_+(0) = \f{2\GT}{\nu L^0_-(0)}\,.\eeq{wh111}
At the same time, as follows from \eq{wh10}
\beq Q(0) = \lim_{k\to \I\infty} (-\I k)Q_+(k)=\lim_{k\to -\I\infty} (\I k)Q_-(k) = C\,.\eeq{112}

In our formulation, the chain-connecting bonds break at $\Gn=0$. So, $Q(0)$ is equal to its critical value $C=Q(0)=Q_*$.  The speed-dependent relation between the external action and the critical strain of the $c_3$-bonds follows as
\beq \GT = \f{\nu}{2} L^0_-(0)Q_*\,,\eeq{113}
and the local fracture energy is
\beq G_0=\f{c_3^2 Q_*^2}{2}= \f{2c_3^2\GT^2}{\nu^2(L^0_-(0))^2}\,.\eeq{113a}

\begin{figure}[h]
\centering \vspace{0mm}
\begin{subfigure}{.5\textwidth}
  \centering
  \includegraphics[width=.7\linewidth]{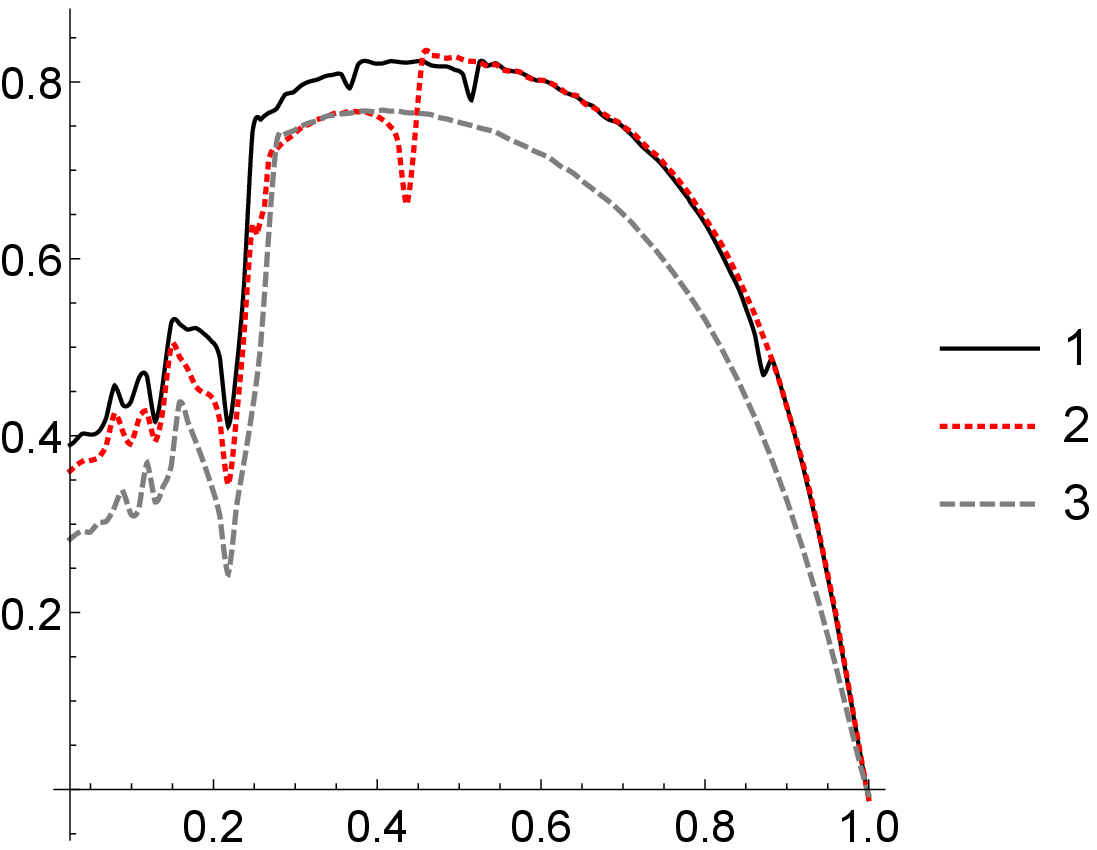}
  \vspace{4mm}
  \caption{$c_1$ = 4 (1), 2 (2), 1 (3)}
  \label{8sub1}
\end{subfigure}%
\begin{subfigure}{.5\textwidth}
  \centering
  \includegraphics[width=.7\linewidth]{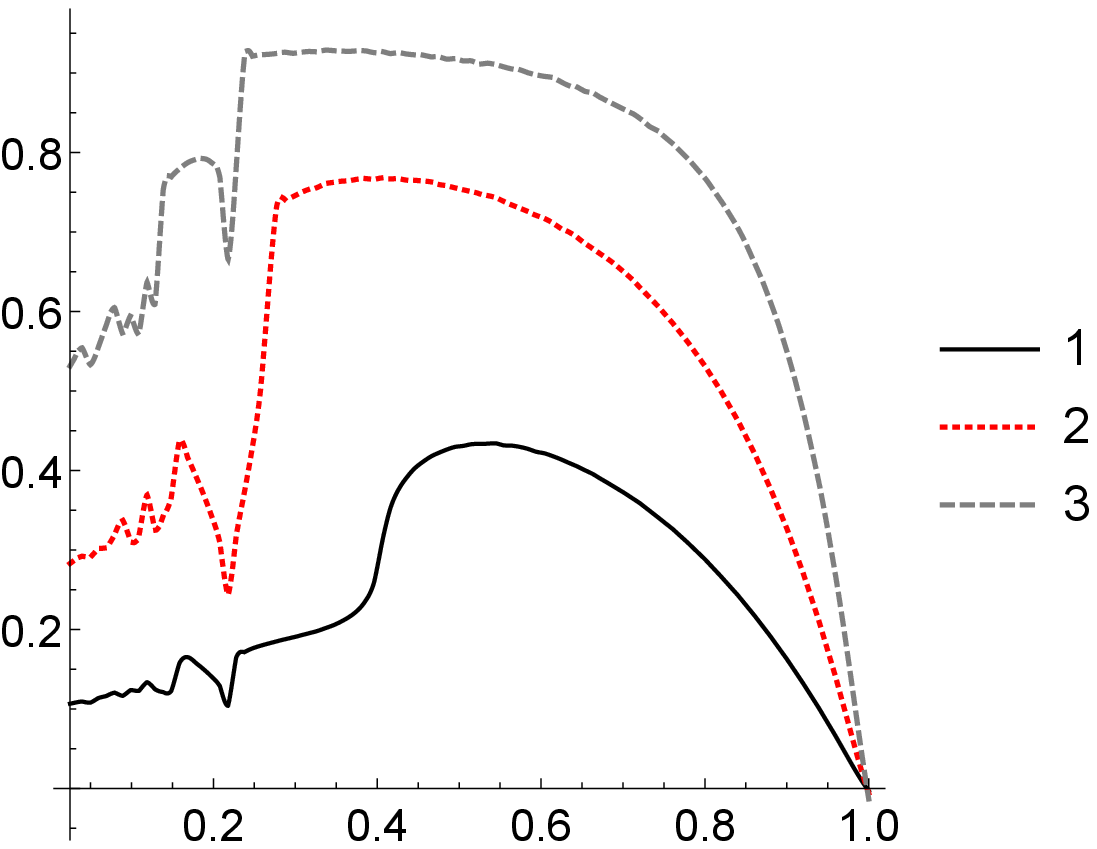}
  \vspace{4mm}
  \caption{$c_3$ = 2 (1), 1 (2), 0.5 (3)}
  \label{8sub2}
\end{subfigure}
\begin{picture}(0,0)
	\put(-212,130){$\f{G_0}{G}$}
	\put( 23,121){$\f{G_0}{G}$}
	\put(-110,27){$v$}
	\put(125,27){$v$}
	\end{picture}
\caption{Dynamic discrete model. Subsonic regime. Energy release ratio for $c_2=c_3=1$ (a) and for $c_1=c_2=1$ (b)}
\label{f10}
\end{figure}

\begin{figure}[h]
\centering \vspace{-0mm}
\begin{subfigure}{.5\textwidth}
  \centering
  \includegraphics[width=.7\linewidth]{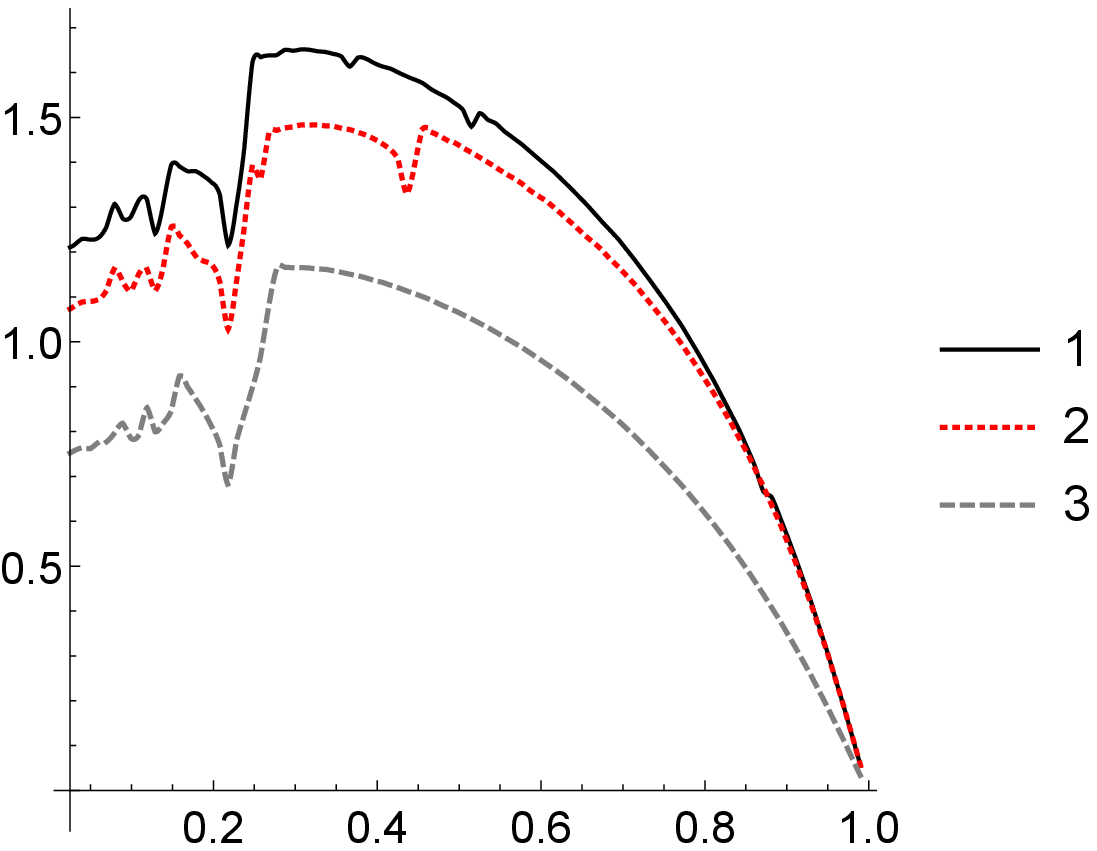}
  \vspace{4mm}
  \caption{$c_1$ = 4 (1), 2 (2), 1 (3)}
  \label{9sub1}
\end{subfigure}%
\begin{subfigure}{.5\textwidth}
  \centering
  \includegraphics[width=.7\linewidth]{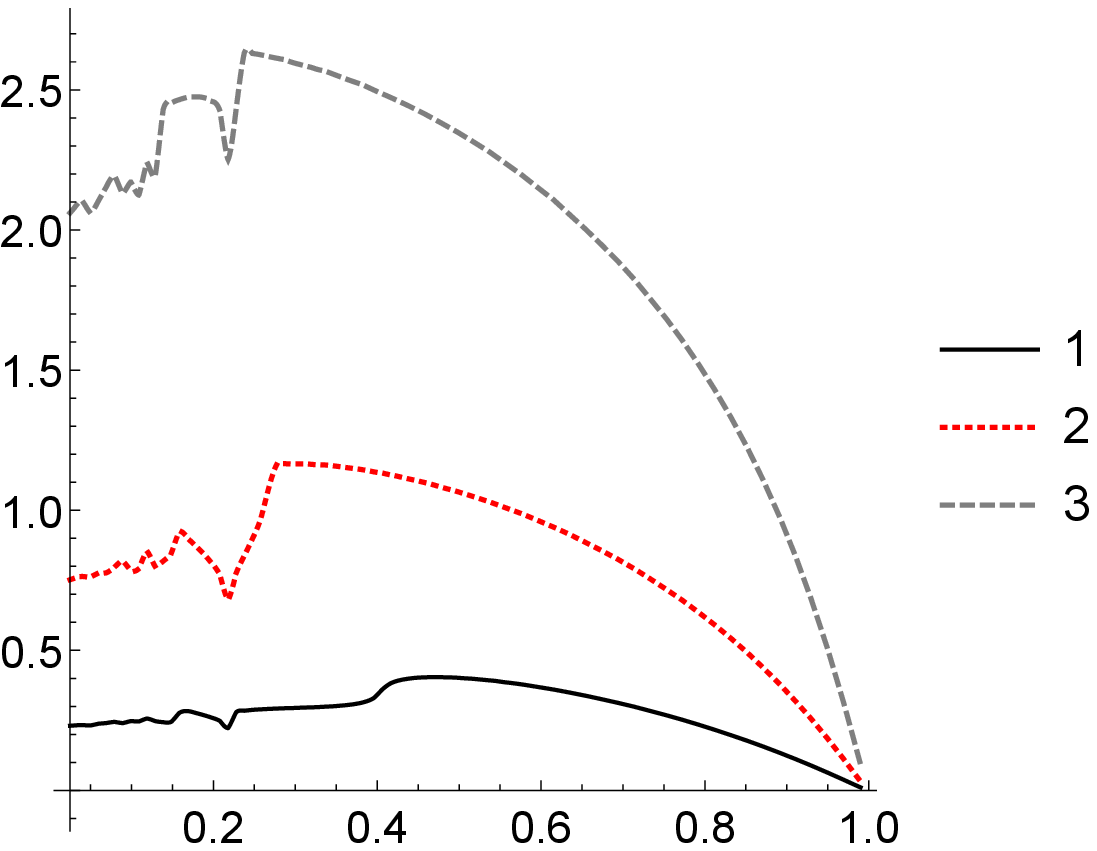}
  \vspace{4mm}
 \caption{$c_3$ = 2 (1), 1 (2), 1/2 (3)}
  \label{9sub2}
\end{subfigure}
\begin{picture}(0,0)
	\put(-210,120){$\f{Q_*}{\Theta}$}
	\put( 20,133){$\f{Q_*}{\Theta}$}
	\put(-110,28){$v$}
	\put(124,28){$v$}
	\end{picture}
\caption{Dynamic discrete model. Subsonic regime. Relation between bond critical strain $Q_*$ and external action $\Theta$ versus the speed $v$ for $c_2=c_3=1$ (a) and for $c_1=c_2=1$ (b).}
\label{f11}
\end{figure}

\begin{figure}[h]
\centering

  \includegraphics[scale=0.5]{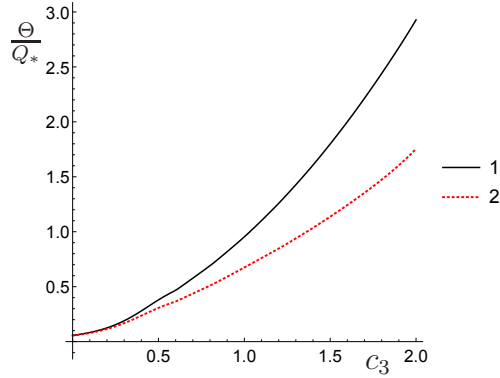}
  \vspace{4mm}
  \caption{Discrete models. Subsonic regime. Quasi-static (1) and minimal dynamic (2) resistance to the splitting  vs. bond stiffness $c_3$; $c_1 = 2, c_2 = 1$.}\label{f13}
  \begin{picture}(0,0)
	\put(-100,179){$\f{\GT}{Q_*}$}
	\put(35,58){$c_3$}		
  \end{picture}
  \label{f12}
\end{figure}

The global energy release obtained for the continuous model \eq{fcr6} is independent of the string connection structure. The only important point is that for $\Gn\to\infty$ the connection leads to uniform strains the same for both strings. So, the `macro-level' (long-wave) energy release rate for the chains is equal to that presented in \eq{fcr6} in terms of $\GT$, namely
\beq G = \f{2(c_1^2-v^2)(c_2^2-v^2)}{c_1^2+c_2^2-2v^2}\GT^2= \f{2c_3^2\GT^2}{\nu^2}\,.\eeq{114}
Thus, the energy ratio is
\beq \f{G_0}{G} =( L_+^0(0))^2 = \exp\gl[\f{2}{\pi}\int_0^\infty \f{\mbox{Arg} L(\xi)}{\xi}\D\xi\gr]\,.\eeq{gg01}
This ratio is plotted in \fig{f10}. The normalized external action $\Theta$  \eq{113} is presented in \fig{f11}. The comparative plots of the quasi-static and the minimal dynamic resistance to the splitting  are shown in \fig{f12}.

\vspace{10mm}

\vspace{0mm}

\subsection{Supersonic regime}
Consider the Wiener-Hopf equation kernel \eq{wh6} in a vicinity of $k=0$
\beq L(k) =\f{h_1h_2 +c_3^2(h_1+h_2)}{h_1h_2} \sim \f{c_3^2(h_1+h_2)}{h_1h_2}\n
\sim \f{c_3^2[(c_1^2 +c_2^2-2v^2)k^2 + 4\I s k v] }{[(c_1^2-v^2)k^2+2\I s k v][(c_2^2 -v^2)k^2 +2\I s k v]} \,.\eeq{wh6ir}
It can be seen that Arg$(L(k))=0$ for the subsonic case and becomes equal to $\pi \,\mbox{sign} k$ in the intersonic range, $c_2<v<c_+= \sqrt{(c_1^2+c_2^2)/2}$, where the energy ratio \eq{gg01} is infinite. Thus, it is the same band gap as in the continuous model considered in \az{ir}.

The argument returns back to zero for the supersonic range $c_+<v<c_1$. Thus, the solution in the latter speed range exists as well as in the subsonic speeds. The condition for $\Gn \to -\infty$ is defined by the long wave asymptote, \az{ir}. Namely, we can use the ratio $\mathbb{P}_2/\CP_1$  from \eq{p1o} (but not the expressions connecting these forces with $Q_*$).
The relation \eq{bc1} looks now as
\beq \GT = \f{v^2-c_+^2}{v^2-c_2^2}\f{\CP_1}{c_1^2}\,.\eeq{bc1a}
From here and \eq{113} we find the force $-$ external action relation
\beq \CP_1 = c_3 c_1^2\sqrt{\f{v^2-c_2^2}{2(v^2-c_+^2)(c_1^2-v^2)}}\, L_-^0(0)Q_*\,.\eeq{irfc3}
Note that this expression differs from that obtained for the continuous model \eq{p1o} only by the factor $L_-^0(0)$. It is plotted  in \fig{f6} $-$ \fig{f7d} together with the related dependencies for the continuum model and the discrete strip in statics.

\subsection{The radiated waves}
The propagating sinusoidal waves emitted by the moving splitting correspond to nonzero real singular points $u^F$ and $w^F$ . Such a point is the wavenumber, and the dispersion relation defines the corresponding frequency $\Go$. Graphically, these parameters coincide with the coordinates of the intersection of the ray  $\Go=k v$ with the dispersion curve.

The sinusoidal wave propagates to the right if $v<c_g$, and vice versa, where $c_g=\D\Go/\D k$ is the group speed. In the case of the transition wave, different dispersion relations correspond to different areas. Recall that this fact imposes an additional condition on the wave radiation. Namely, the waves with $v<c_g$ propagate to the right if they correspond to the dispersion relations valid for $\Gn>0$, and vice versa. Such `active' points are marked by small circles on the dispersion relations shown in \fig{f8} and \fig{f9}.

From \eq{wh2} and \eq{wh10aa} with $C=Q_*$ we get
\beq u^F(k) = \f{c_3^2Q_*}{(\nu-\I k)L^0_+(k)h_1(k)}\,,~~~ w^F(k) =- \f{c_3^2Q_*}{(\nu-\I k)L^0_+(k)h_2(k)}\,.\eeq{rw1}
Let $k_0$ be a simple pole of one of these functions. If and only if it is located on the real $k$-axis, the corresponding residual defines the propagating wave.More precisely, if the pole is $k=k_0 + \I 0$ then the wave
\beq u_0(\Gn) = u^0\E^{-\I k_0\Gn}\,,~~~\mbox{or}~~~w_0(\Gn) = w^0\E^{-\I k_0\Gn}\,,\n u^0=  \lim_{k\to k_0}(k-k_0)u^F(k_0)\,,~~~w^0= \lim_{k\to k_0}(k-k_0)w^F(k_0)\eeq{rw2}
propagates from the transition point to the left. Otherwise, if $k=k_0-\I 0$
\beq  u_0(\Gn) = -u^0\E^{-\I k_0\Gn}\,,~~~\mbox{or}~~~w_0(\Gn) = - w^0\E^{-\I k_0\Gn}\,,\eeq{rw3}
and the wave propagates to the right. By the symmetry, for $k_0\pm \I0$ there exists the same pole at $k= - k_0 \pm \I 0$.

The period-averaged energy density of such a wave is
\beq \CE=\Go_0^2|u^0|^2~~\mbox{or}~~\CE=\Go_0^2|w^0|^2~~~(k_0=0\pm\I 0)\,.\eeq{rw4a}
With the account of the similar wave related to the symmetric pole $ k=- k_0\ne 0$ the total energy becomes four times greater. Here we take into account the fact that the averaged kinetic and potential energies of linear waves are equal (in this connection, see Slepyan (2015)). The energy flux from the transition point (as the energy per unite time) is
\beq N = \CE |c_g(k_0) - v|\,.\eeq{rw5}
The total energy release rate associated with the waves (per unit length of the splitting) defined in \az{tdpds} is the difference between its global and local values.

\subsection{Distribution of the connecting bond strain and admissibility of the solutions}\label{aots}
The solutions obtained are admissible if the strain  $Q$ first reaches the critical value $Q_*$ at $\Gn=0$ but not ahead of the transition point (Marder and Gross, 1995). Otherwise,  the steady-state solution does not hold. Recall that in the latter case, other established regimes can form. The clustering and forerunning modes of the chain-strip fracture and a flexural beam detachment were disclosed in Mishuris et al. (2009) and Slepyan et al.(2015), respectively, both under a sinusoidal wave action.

To inspect the admissibility we now return to the solution in \eq{wh10aa}.
\beq Q_+(k) =\f{C}{(\nu-\I k)L^0_+(k)}\eeq{wh10aaa}
with $C=Q_*$. So we have the double integral, the Cauchy type integral for the factorization and the inverse Fourier transform.
In order to obtain a good-accuracy result by a numerical evaluation of this analytical representation, we first make further normalization of the kernel $L(k)$ by developing the function $S(k)$ \eq{wh8} to incorporate all poles and zeros of $L$ on the real $k$-axis.
\beq S(k) = S_+S_-\,,\n S_+= \f{\nu-\I (k+\I 0)}{(0-\I k)}\prod_{i}\f{k_i^2 -(k+\I 0)^2}{p_i^2 -(k+\I 0)^2}\,,~~~S_-= \f{\nu+\I (k-\I 0)}{(0+\I k)}\prod_{j}\f{k_j^2 -(k-\I 0)^2}{p_j^2 -(k-\I 0)^2}\,,\eeq{sd1}
where $k=k_i \, (k_j)$ and $k=p_i \, (p_j)$ are (positive) zeros of the numerator and denominator of $L(k)$, \res. The number of the zeros increases as the transition speed $v$ decreases. For the strip where $c_1/c_2=2, c_3/c_2=1$ considered below, the are no such zeros in $S_+$ if $v> v_*=0.4497495521 c_2$. In this speed range, there is no sinusoidal waves propagating ahead of the transition point, and only one such a wave exists in $v_*< v < v_{**}=0.2654575261 c_2$.

In these terms
\beq L^0(k) = \f{L(k)}{S(k)} = L_+L_-\,,~~~L_\pm = L_\pm^0 S_\pm\eeq{sd2}
and for $\Gn\ge 0$
\beq Q(\Gn) = \f{Q_*}{2\pi}\inti \f{1}{(0-\I k)L_+(k)}\E^{-\I k \Gn}\D k\,.\eeq{sd3}
At the speeds $v<v_{*}$ there exist sinusoidal waves at $\Gn>0$ and the corresponding poles in the integrant. For this speed region, we represent the integral in \eq{sd3} as a sum of half-residues and the Cauchy principal value, which is calculated numerically as well as the factorization integral in the first line in \eq{wh9}.We consider both these ranges, $v>v_{**}$, noting the steady-state regime at lower speeds is at least unstable, and the crack in a lattice cannot propagate steady at low speeds\footnote{The function $Q(\Gn)$ corresponding to the one-sided Fourier transform $Q_+(k)$ is zero at $\Gn<0$; hence, it has a jump discontinuity at $\Gn =0$. In turn, this results in a weak conversion of the inverse Fourier transform, which does not allow obtaining a correct result for a vicinity of this point. We improve the conversion separating a term having a simple analytical expression.} (in this connection, see \fig{f10}a).

\begin{figure}[h]
\centering \vspace{0mm}
\begin{subfigure}{.5\textwidth}
  \centering
  \includegraphics[width=.7\linewidth]{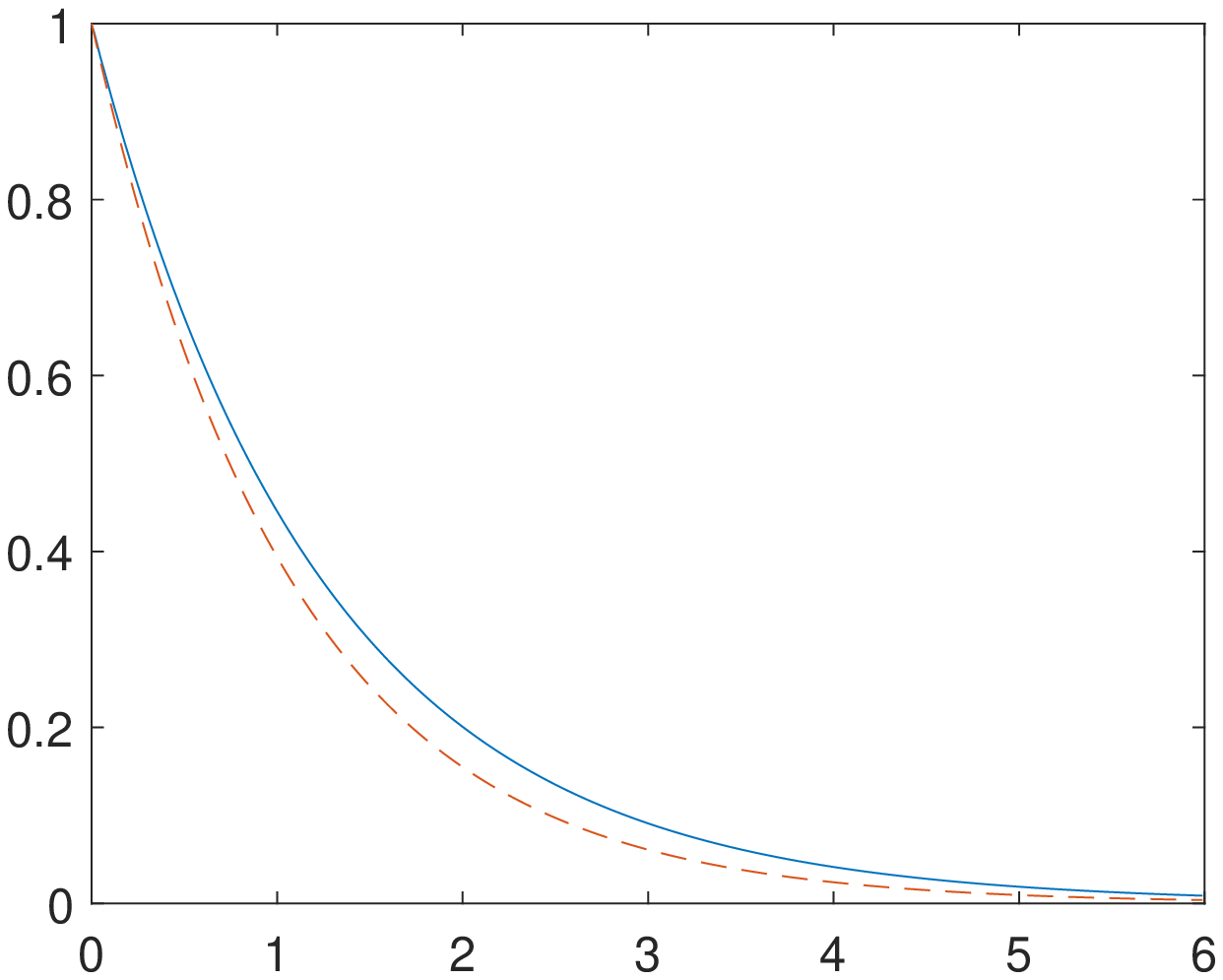}
  \vspace{4mm}
  \caption{The supersonic speed $v=1.8 c_2> c_+$.}
  \label{14sub1}
\end{subfigure}%
\begin{subfigure}{.5\textwidth}
  \centering
  \includegraphics[width=.7\linewidth]{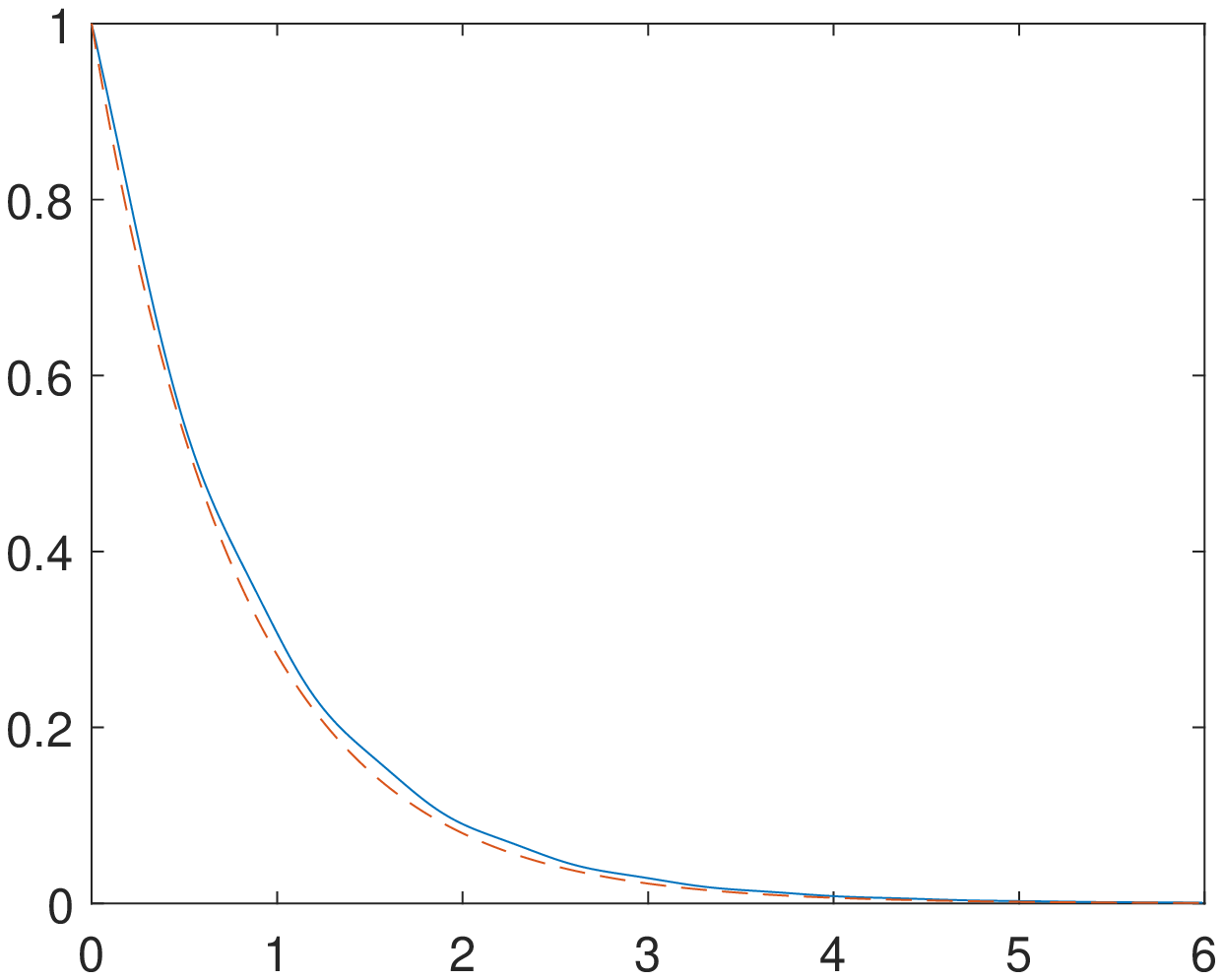}
  \vspace{4mm}
  \caption{The subsonic speed $v=0.5 c_2> v_*$.}
  \label{14sub2}
\end{subfigure}
\begin{picture}(0,0)
	\put(-203,113){$\f{Q}{Q_*}$}
	\put( 32,113){$\f{Q}{Q_*}$}
	\put(-110,32){$\Gn$}
	\put(125,32){$\Gn$}
	\end{picture}
\caption{The normalised connecting bond strain $Q(\Gn)/Q_*$ ahead of the transition point for the discrete (the solid curve) and related continuous (the dashed curve) systems; $c_3=c_2$. In the latter case, $Q(\Gn)/Q_*=\exp(-\nu\Gn)$. There is no sinusoidal wave in this region.}

\label{a1}
\end{figure}

\begin{figure}[h]
\centering \vspace{0mm}
\begin{subfigure}{.5\textwidth}
  \centering
  \includegraphics[width=.7\linewidth]{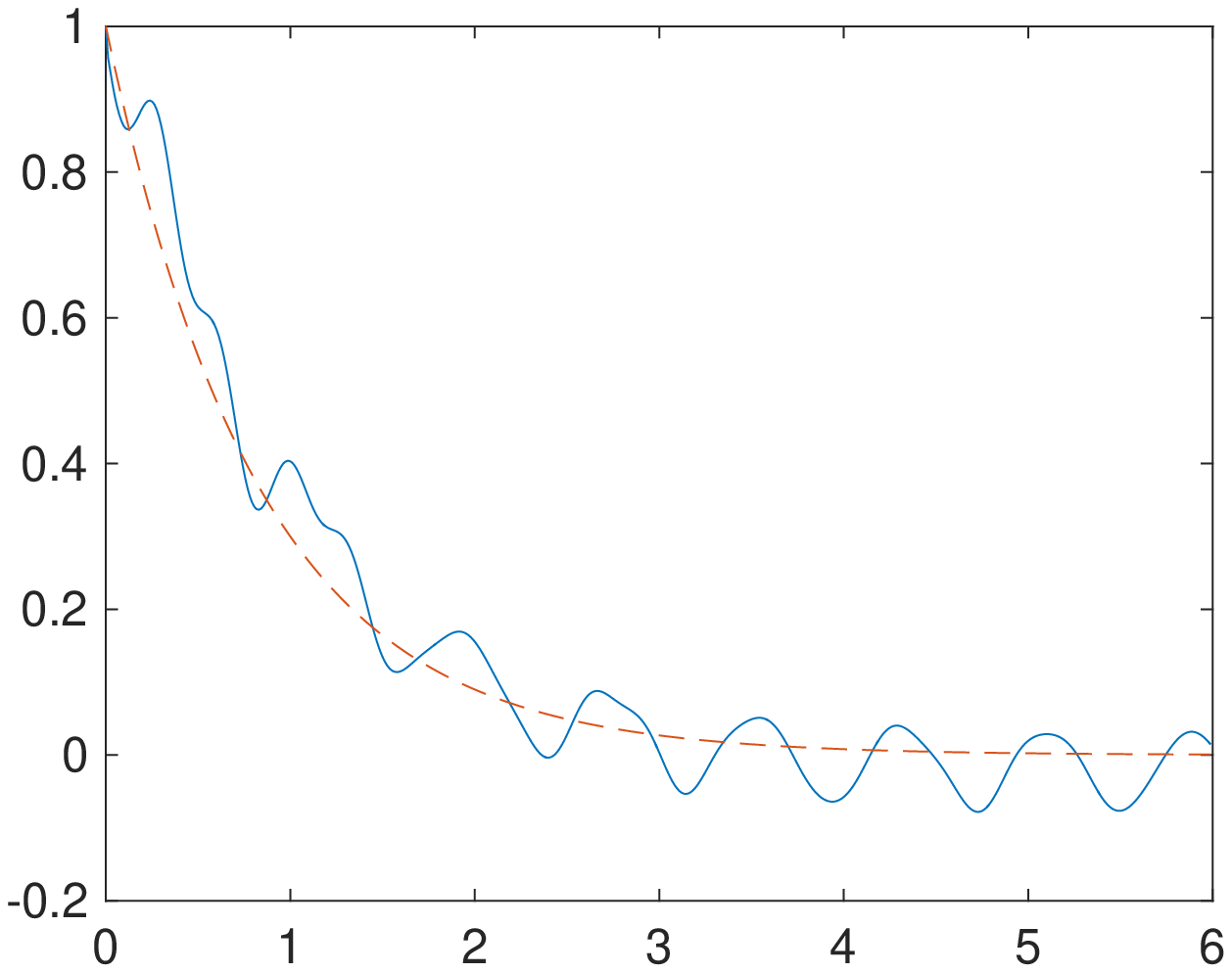}
  \vspace{4mm}
  \caption{The subsonic speed $v=0.4 c_2<v_*$.}
  \label{15sub1}
\end{subfigure}%
\begin{subfigure}{.5\textwidth}
  \centering
  \includegraphics[width=.7\linewidth]{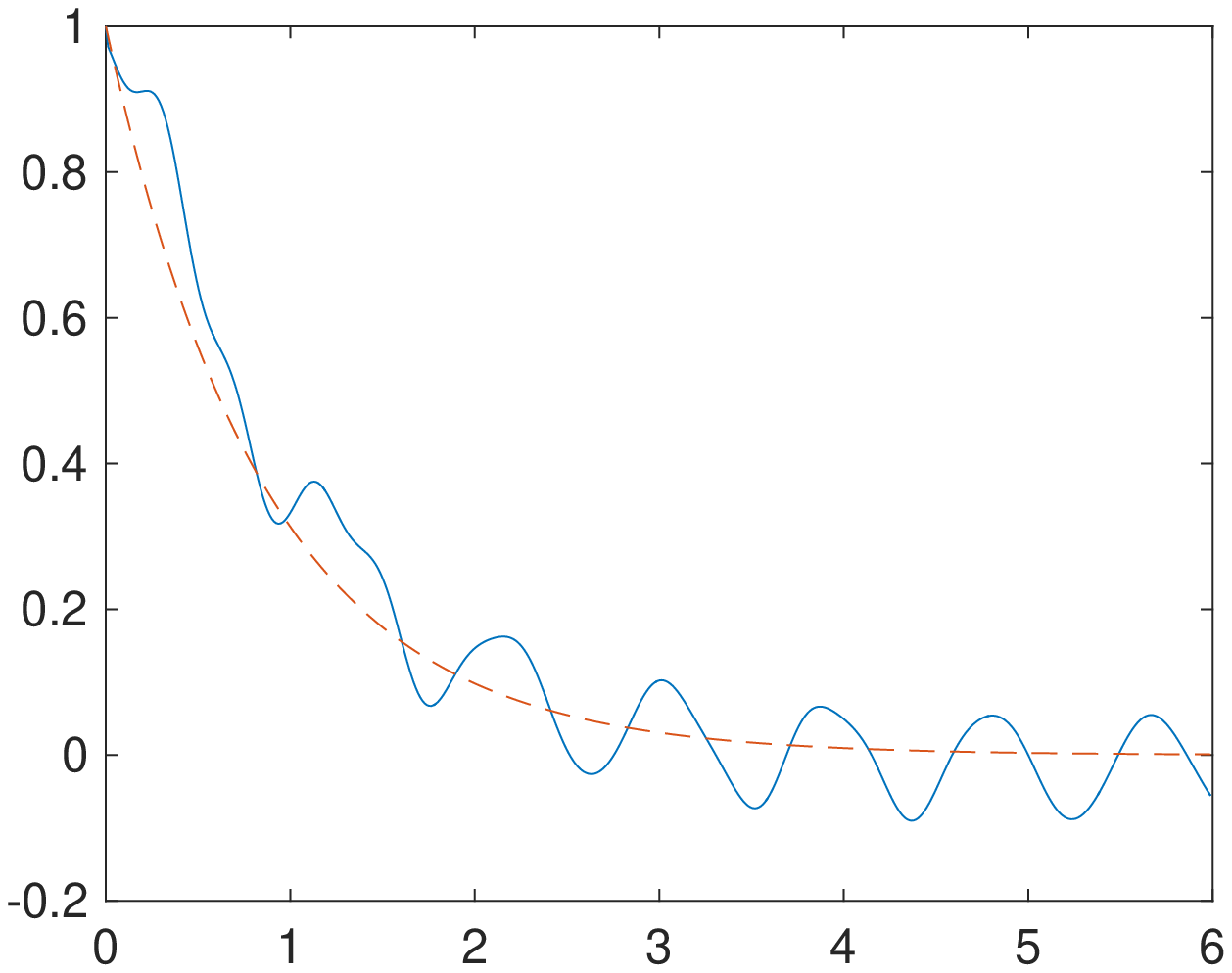}
  \vspace{4mm}
  \caption{The subsonic speed $v=0.29 c_2<v_*$.}
  \label{15sub2}
\end{subfigure}
\begin{picture}(0,0)
	\put(-204,118){$\f{Q}{Q_*}$}
	\put( 31,118){$\f{Q}{Q_*}$}
	\put(-110,33){$\Gn$}
	\put(125,33){$\Gn$}
	\end{picture}
\caption{The normalised connecting bond strain $Q(\Gn)/Q_*$ ahead of the transition point for the discrete (the solid curve) and related continuous (the dashed curve) systems; $c_3=c_2$. In the latter case, $Q(\Gn)/Q_*=\exp(-\nu\Gn)$. There exists a single sinusoidal wave in this region.}
\label{a2}
\end{figure}

The calculation results are shown in \fig{a1} and  \fig{a2}. For $v>v_*$, the function $Q(\Gn)/Q_*$  almost coincides with $\exp(-\nu\Gn)$ corresponding to the continuum model. So, at these speeds discreteness weakly affects the strain in the region ahead of the transition point. Note, however, that there exist sinusoidal waves propagating to the left along the separated chains.

For $v_*<v<v_{**}$, \fig{a2}, the dependence is not monotonic (because of the presence of a sinusoidal wave) but also reaches maximum only at $\Gn=0$. Thus, the considered solutions are admissible for both the subsonic, $v_{**}<v<c_2$, and supersonic, $c_+<v<c_1$, speed regions.

\section{Conclusions}
As compared with the identical-chain strip, the dissimilar-chain system represents a totally different waveguide. It is characterized by three long wave speeds, $c_1$ and $c_2$ for the separate chains and $c_+=\sqrt{(c_1^2+c_2^2)/2}$  for the strip of connected chains.  Accordingly, there exist three ranges, the subsonic $(0, c_2)$, intersonic $(c_2, c_+)$  and supersonic $(c_+,c_1)$.

We find that the splitting can propagate in two of these speed ranges, the subsonic and supersonic, whereas the steady splitting in the intersonic regime is impossible. We also find that in the case of considerable difference in the chain stiffness, the lowest dynamic threshold corresponds to the supersonic regime.

Thus, in such a composite structure, the `splitting wave' can propagate supersonically. It can be noted that under some conditions, this regime can appear even in a uniform structure. Such an extraordinary mode can exist in any case where the required energy can be delivered to the crack tip, in particular, in a stressed lattice ( Slepyan, 2002, pp. 405-406, Buehler and Gao, 2006,  and Ayzenberg-Stepanenko et al., 2014 ). The peculiarity of the regime under consideration is that the energy transfers from the remote forces, and {\em the supersonic energy delivery channel initially absent is opening with the splitting}. It is a nontrivial manifestation of the dissimilarity.

Like a crack in a brittle body, the splitting of the chain strip cannot grow quasi-statically. The minimal resistance to the transition in the subsonic speed range is below that in the quasi-static regime. The new fact is that the splitting can grow at supersonic speeds.

The same speed ranges, permitted and forbidden, are inherent for the continuum approximation of the discrete model. Although no sinusoidal wave radiates in the continuous model, it represents the `macrolevel' framework for the chain strip. Thereby, the continuous model defines the total energy release for both the quasi-static and dynamic regimes.

For not too low speeds, where no more than one sinusoidal wave penetrates the area in front of the moving splitting point, we present the strain distribution ahead of the transition point for both the supersonic and subsonic speeds.  In these speed regions, the strain decreases with the distance from the transition point (possibly, with some oscillations), where it reaches the global maximum. Thus, {\em  the solutions satisfy the Marder's admissibility condition}.

We find that in the supersonic and a part of the subsonic regimes, where there are no sinusoidal waves propagated in front of the splitting point, the strain distribution almost coincides with that for the continuum model. It means that at these speeds, discreteness weakly affects the dynamic field ahead of the transition point.

We also find the force $-$ speed relations, the local-to-global energy release ratio and parameters of the sinusoidal waves radiated from the transition point.

\vspace{20mm}
\vskip 18pt
\begin{center}
{\bf  References}
\end{center}
\vskip 3pt

\inh Alberts, B., Johnson, A., Lewis, J., Raff, M., Roberts, K., Walter, P.  2002. Molecular Biology of the Cell. Garland Science. ISBN 0-8153-3218-1.

\inh Ayzenberg-Stepanenko, M.V., Mishuris, G.S., and Slepyan, L.I., 2014. Brittle fracture in a periodic structure with internal potential energy. Spontaneous crack propagation.  Proc. R. Soc. A 2014 470, 20140121.

\inh  Banks-Sills, L., 2015.  Interface fracture mechanics: theory and experiment. Int J Fract 191, 131–146.

\inh Buehler, M.J., and . Gao, H.J, 2006. Modeling Dynamic Fracture Using Large-Scale Atomistic Simulations. In: A. Shukla (ed.), Dynamic Fracture Mechanics, World Scientific, 1-68.

\inh Buehler, M.J., 2008. Atomistic modeling of materials failure. Springer Science \& Business Media.

\inh Eshelby, J.D., 1951. The Force on an Elastic Singularity. Phil Trans Roy Soc (London) A224, 87-112

\inh Eshelby, J.D., 1956. The Continuum Theory of Lattice Defects. In: Seitz F. and Turnbull D. (Eds) Progress in Solid State Phys 3. Academic, New York, 79-144.

\inh Griffith, A.A., 1920. The Phenomena of Rupture and Flow in Solids. Phil Trans Roy Soc (London) A221, 162-198.

\inh Kunin, I.A., 1975. The theory of elastic media with microstructure. Nauka (in Russian).

\inh Kunin, I.A., 1982. Elastic Media with Microstructure. Springer-Verlag.

\inh Kunin, I.A., 1983. Elastic Media with Microstructure II. Springer-Verlag.

\inh Langlet, A., Safont, O. and Renard, J., 2012. The response of infinite strings and beams to an initially applied moving force: Analytical solution. Journal of Vibration and Acoustics, 134(4), 041005.

\inh Liu, W.K., Karpov, E.G., and Park, H.S., 2006. Nano Mechanics and Materials: Theory, Multiscale Methods and Applications. John Wiley \& Sons.

\inh Marder, M., and Gross, S., 1995. Origin of crack tip instabilities. J of the Mech. Phys. Solids 43, 1 - 48.

\inh Mishuris, G.S., Movchan, A.B., Slepyan, L.I., 2008. Dynamical
extraction of a single chain from a discrete lattice. J. Mech. Phys. Solids 56, 487-495.

\inh Mishuris, G.S., Movchan, A.B., Slepyan, L.I., 2009. Localised knife
waves in a structured interface. J. Mech. Phys. Solids 57, 1958-1979.

\inh Mishuris, G.S., Movchan, A.B., and Bigoni, D., 2012. Dynamics of a fault steadily propagating within a structural interface. Multiscale Model. Dimul. Society for Industrial and Applied Mathematics, 10(3), 936–953.

\inh Mishuris, G.S., and Slepyan, L.I., 2014. Brittle fracture in a periodic structure with internal potential energy.  Proc. R. Soc. A., 470:20130821 Published 19 February 2014.

\inh 	Novozhilov, V., 1969a.  On a necessary and sufficient criterion for brittle strength. Journal of Applied Mathematics and Mechanics 33(2), 201-210 (PMM  33(2), 212-222).

\inh Novozhilov, V., 1969b. On the foundations of a theory of equilibrium cracks in elastic solids. Journal of Applied Mathematics and Mechanics 33(5), 777-790 (PMM vol. 33(5), 797–812).

\inh Slepyan, L.I., 1981. Dynamics of a crack in a lattice. Sov.
Phys. Dokl., 26, 538-540.

\inh Slepyan, L.I., and Troyankina, L.V., 1984. Fracture Wave in a
Chain Structure. J. Appl. Mech. Techn. Phys., 25, No 6, 921-927.

\inh Slepyan, L.I., 2002. Models and Phenomena in Fracture
Mechanics. Springer, Berlin.

\inh Slepyan, L.I., Movchan, A.B., Mishuris, G.S., 2010 (2009 online).  Crack in a lattice waveguide. Int J Fract 162, 91-106.
IUTAM Symposium on Dynamic Fracture (K. Ravi-Chandar and T.J. Vogler, Edrs), 2010, 91-106. Springer, Dordrecht.

\inh Slepyan, L.I., Ayzenberg-Stepanenko, M.V., and Mishuris, G.S., 2015. Forerunning mode transition in a continuous waveguide. J. Mech. Phys. Solids 78, 32-45.

\inh Slepyan, L.I., 2015. On the energy partition in oscillations and waves.  Proc. R. Soc. A 471: 20140838.

\inh Thomson, R., Hsieh, C., and Rana, V., 1971. Lattice Trapping of Fracture Cracks. J. Applied Physics 42(8), 3154-3160.

\end{document}